\documentclass[preprint,12pt]{elsarticle}

\usepackage{amssymb}
\usepackage{color}
\definecolor{pblue}{rgb}{0.13,0.13,1}
\definecolor{pgreen}{rgb}{0,0.5,0}
\definecolor{pred}{rgb}{0.9,0,0}
\definecolor{pgrey}{rgb}{0.46,0.45,0.48}
\definecolor{dgreen}{rgb}{0,0.35,0}
\usepackage{graphicx,amsmath,amsfonts}
\usepackage{xcolor}
\usepackage{transparent}
\usepackage{hyperref}
\usepackage{booktabs}
\usepackage[final]{changes}
\usepackage{lipsum}

\usepackage{listings}
\usepackage{enumitem}
\newlist{questions}{enumerate}{2}
\setlist[questions,1]{label=RQ\arabic*.,ref=RQ\arabic*}
\setlist[questions,2]{label=(\alph*),ref=\thequestionsi(\alph*)}

\usepackage{caption}
\usepackage{subcaption}

\graphicspath{{figures/}}

\newcommand{\systime}{\texttt{systime}}
\newcommand{\perf}{\texttt{perf}}
\newcommand{\rd}[1]{\textcolor{red}{#1}}

\journal{Journal of Computer Languages}

\begin{document}

\begin{frontmatter}

\title{\replaced{Investigating}{Predicting} the Energy Consumption of C++ \added{and Java} Solutions Mined from a Programming Contest Site}

\author[inst1]{Sérgio Queiroz de Medeiros}

\affiliation[inst1]{organization={School of Sciences and Technology, Federal University of Rio Grande do Norte},
            city={Natal},
            state={RN},
            country={Brazil}}

\author[inst1]{Marcelo Borges Nogueira}
\author[inst1]{Gustavo Quezado}

\begin{abstract}
The current concern about global warming has led to an increasing interest in the energy efficiency of computer applications. Assuming power is constant, the general trend is that faster programs consume less energy, thus optimizing a program for speed would also improve its energy efficiency. 

We investigate this tendency in a set of C++ \added{and Java} solutions mined from Code Submission Evaluation System (CSES), a popular programming competition site, where each solution must give the correct answer under a given time limit. In such context, we can consider that all correct solutions for a problem were written with a speed concern, but not with energy efficiency in mind.

We selected 15 problems from CSES and for each of them\deleted{,} we mined \added{at least} 30 C++ \added{and Java} solutions, evaluating time and energy efficiency of each solution in \added{at least} two different machines. In our scenario, where there is a great diversity of programming styles, execution speed, and memory usage, we could confirm the general trend: faster programs consume less energy. Moreover, we were able to use ordinary least squares to \replaced{fit a linear function}{predict}, with good precision, \replaced{that relates}{the} energy consumption of a program based on its execution time, as well as to automatically identify programs with abnormal energy consumption. A manual analysis of these programs revealed that often \replaced{they perform a different amount of allocation and deallocation operations}{their memory usage is different} when compared to programs with similar execution times.

We also calculated the energy consumption profile of sets of random \added{C++} solutions for these 15 CSES 
problems, and we tried to associate each set with its corresponding CSES problem by using the
energy consumption profiles previously computed for each one of them. By using this approach, we 
could restrict, for each set of random \added{C++} solutions, the classification task to a subset of 7 CSES 
problems, a reduction of more than 50\% in the search space.
\end{abstract}


\begin{highlights}

\item We present a methodology to compute the energy consumption profile of a computational problem in a given machine based on mined C++ \added{and Java} solutions for it.

\item \added{We check the effect of several factors, such as optimization
flags and the availability of multiple cores,
on the energy consumption profile computed of a problem.}

\item We \replaced{confirm the strong linear correlation between execution time and}{predict the} energy consumption of C++ \added{and Java} programs, for a given computational problem in a given machine.

\item By using our methodology, we could automatically identify \added{C++} programs with abnormal energy consumption.

\item We could give a good estimate of which one of 15 problems a given set of randomly chosen C++ solutions is aimed to solve by associating its energy consumption profile with one of the problems.
\end{highlights}


\end{frontmatter}

\section{Introduction}

Computing has an ever-increasing impact on global energy consumption. By 2012, almost 5 percent of the world's electrical energy was consumed by computers and other related devices~\cite{gelenbe2015ictimpact}. The current concern with the global warming effects has also increased the concern about energy consumption when developing software~\cite{pinto2017energyconcern,marantos2022bringing}. 

The energy consumed by an application running in a computer depends on the computer's power and also on the application execution time. Overall, we could diminish the energy consumption of an application by using one of the following \deleted{two} strategies:
\begin{itemize}
    \item \textbf{Power saving:} We could use Dynamic Voltage and Frequency Scaling (DVFS), a technique that allows for dynamically adjusting the operating frequency and voltage, to consume less power while a program runs. By using this, our program would run slower, but we could get energy savings due to lower power used.

    \item \textbf{Faster execution time:} If we were able to reduce the execution time of a program, assuming that the power consumption does not increase considerably, we could get energy savings. 

    \item \textbf{\added{Avoid wasteful computations:}} \added{For some applications, operations such as
    busy waits and transaction retries sometimes can lead to an improved performance,
    but at the cost of extra energy being consumed.}
\end{itemize}

High-level energy models~\cite{yuki2014folklore} indicate that, due to factors such as leakage power, we should not expect energy savings in current computers from the usage of DVFS. Thus, in practice, we can often assume that power is constant during the execution of a program, and we can express the energy consumption of a program with
the following simple equation:
\[
Energy \;=\; Power \times Time
\]
\replaced{Thus, disregarding operations such as transaction retries,}{Based on it}, a program with a faster execution time should consume less energy when compared with a slower one.

Although we can often assume that the CPU power is constant, there are still other factors that can impact the general energy consumption of a program, such as memory usage and thread synchronization~\cite{pinto2017forkjoin,pinto2014thread}. In our scenario, where we analyze single-threaded programs, only memory usage should have an influence on energy consumption.

Previous studies have investigated, for a given benchmark, the energy consumption of solutions implemented in different programming languages~\cite{couto2017sblp,pereira2017energy,pereira2021ranking}.
Overall, it was observed a very strong correlation between the execution time of a program and its energy consumption and a strong correlation between the total memory usage of a program and its Dynamic Random Access Memory (DRAM) energy consumption.

Inspired by these studies, we performed a study to investigate the energy consumption of different C++ \added{and Java} solutions for a given problem. 
The first goal of our study is to evaluate the claim of previous research that energy consumption correlates strongly with execution time. In case this is true, we also want to investigate the possibility of characterizing the computational power of a machine, in the context of a given problem, given a set of programs that solve it \deleted{, and then use this to predict the energy consumption of a new solution based on its execution time}.

In order to construct our data set, we mined C++ \added{and Java} solutions from Code Submission Evaluation System (CSES), a popular programming competition site. The CSES archive has 300 problems, and each problem has a series of test cases. A test case consists of a given input and its corresponding output. In order to be considered correct, the solution of a user for a given problem must give for each of its test cases, under a given time limit, the expected answer. In the case of CSES, usually, a program running in the CSES server must produce the correct answer for a test case in at most one second. In such context, we can consider that all correct solutions for a problem were written with a speed concern, but not with energy efficiency in mind.

Initially, we selected 15 problems from CSES and for each one of them we mined 30 C++ solutions, trying to select the faster and the slower ones, as well as some randomly chosen solutions. \added{We also built a C++ and a Java data set which have only random picked solutions for each problem.} Then, we evaluated the time and energy efficiency of each solution in \added{at least} two different machines. The construction of this benchmark, which is publicly available, is also a contribution of our work.

We also conducted a preliminary investigation on how well the energy consumption profile of a given problem is a good discriminator of this problem. The classification of the algorithm implemented
by a program is relevant for redundancy elimination, plagiarism detection and malware identification, for example~\cite{classification2023cgo,malware2013}. Usually, the algorithm classification problem consists
of finding which algorithm a program implements among a given finite set of algorithms.

In our investigation, given a set of \added{C++} programs from CSES that solve a particular problem, we tried to indicate, from a finite set of problems, a restricted subset which includes the particular problem the set of programs are aimed to solve. For this, we characterized the energy consumption profile of each one of the 15 problems (in a fixed machine) and compare it with the set of problems we want to classify.



In this context, we can summarize our research questions as follows:
\begin{questions}

\item \textit{\replaced{By using a small set of C++ and Java programs, aimed to solve a given problem, can we get a linear function that fits the corresponding energy consumption and execution time data points}{Can we confirm that, for a diverse set of mined C++ solutions, there is a strong correlation between energy consumption and execution time}?} \label{rq:question1}

\item \textit{\replaced{By computing an estimate of the expected energy consumption of a program, given its execution time, can help us to identify C++ programs with an abnormal energy consumption}{By using a small set of C++ programs, aimed to solve a given problem, can we get a linear function that predicts energy consumption given execution time}?} \label{rq:question2}

\item \textit{Is it possible to identify what problem a set of \added{C++} solutions intends to solve based on its energy consumption profile?} \label{rq:question3}

\end{questions}

In our scenario, where there is a great diversity of programming styles, execution speed, and memory usage \added{(including the variability on its allocation and deallocation)}, we could confirm the general trend, where faster programs consume less energy. Moreover, using the least square method, we were able to \replaced{fit a linear function }{predict}, with good precision, that relates energy consumption of a program \replaced{and}{based on} its execution time. A manual analysis of the programs with unexpected energy consumption revealed that often
\replaced{they perform a different amount of allocation and deallocation operations}
{their memory usage is different} when compared to programs with a similar execution time.

Besides helping to confirm that faster programs usually are more energy-efficient than slower ones, our results seem to indicate that our approach to \replaced{estimate}{predicting} the energy consumption of a program can help to automatically identify programs with abnormal energy consumption, considering the energy consumption of the entire \added{CPU} socket (see discussion about the methodology in Section~\ref{sec:method}).

Regarding the classification task, we conclude that the classification of a problem given its energy consumption profile is a promising method and deserves further investigation. By using this approach --- and using just a few solutions for training and for the test set --- we could restrict, for each set of random solutions, the classification task to a subset of 7 CSES problems (instead of the 15 original ones), a reduction of more than 50\% in the search space.

In summary, the main contributions of our work are:

\begin{itemize}
\item Publicly available data (two benchmarks with 450 C++ programs each\added{, one benchmark with 450 Java programs,} the analysis script, and the test results) supporting that energy consumption increases linearly with execution
time~\cite{medeiros2024dataset}.

\item An approach to automatically \replaced{identify}{predict} outliers \deleted{programs}, whose energy efficiency is better or worse than \deleted{similar} solutions \replaced{with a similar execution time}{(execution time)}. 

\item An analysis that programs with poor memory management, presenting memory leaks and/or heavy use of memory allocation/frees, tend to perform \replaced{worse}{worst} energetically.

\item A classification method of problems based on its energy consumption profile.
\end{itemize}

This paper is an extended version of a previous work~\cite{energy2023sblp}.
In particular, we performed all the experiments again, since we updated
the operating system of one of the machines and improved our measurement
methodology. Thus, although we restate the conclusions of our previous
paper, the discussion presented here is based on new experimental data.

\added{In this extended version, we also analyze the energy consumption
of Java solutions, and we added a new data set of C++ solutions. Both of
these two new data sets are formed only by randomly accepted solutions
for selected CSES problems.}

Regarding the methodology, we updated the measuring framework
to use the CPU time instead of the wall-clock time when measuring
\replaced{the running time of}{C++} solutions.
\added{In addition, we also used another measurement framework based on
the \perf\, Linux tool, and we also performed the experiments on a
new machine. Finally, we also investigated the effect of using a single
core during our experiments.}
\deleted{Although the results we present are similar, this change allowed us to catch some errors,
as we could see when there was a meaningful difference between both times.}Section~\ref{sec:energypmeasurements} discusses these issues in more detail.

Moreover, \deleted{we constructed a new data set, formed only by 
randomly selected solutions, and based on it} we investigated
the automatic classification of problems based on its energy
consumption profile. Our preliminary results indicate that
such a profile can help to reduce the group of possible problems
that a given computational solution is trying to solve.

\textbf{The rest of this paper is organized as follows:} next section introduces CSES, talks about its problem set, and presents programming strategies often used in programming competitions. Section~\ref{sec:method} explains our methodology to construct and analyze our data set of C++ \added{and Java} programs and to measure the energy consumption of these programs, while Section~\ref{sec:results} presents our results. Section~\ref{sec:related} discusses other works related to the energy consumption of computer applications. Finally, Section~\ref{sec:conc} presents our conclusions and discusses further research directions.

\section{Programming Competition Sites}
\label{sec:cses}

CSES\footnote{\url{https://cses.fi/}} is a website used by programmers to practice for programming competitions. Programming competition problems are often used when teaching disciplines focused on data structures and algorithm techniques~\cite{skiena1998algorithm}, and are also used in coding interviews by companies such as Google and Amazon~\cite{mongan2012programming,mcdowell2015cracking}.
Typical users of sites like CSES are programmers who want to improve their programming and algorithm skills, or who are studying for a job interview.

In a programming competition, there is a set of problems and each problem contains a set of test cases. A test case consists of an input and its corresponding output, the expected answer that a solution for that problem must produce. 

When a user submits a solution to a problem, it is tested against each problem test case, and it is considered correct only when it passes all of them. Moreover, when executing a test case there are also time and memory limits that must be respected. In the case of CSES, usually, a solution for a problem must give the correct answer for a test case in at most one second and must use at most 512 MB of memory.

\begin{figure}[t]
  \begin{center}
    \leavevmode
    \includegraphics[width=0.8\textwidth]{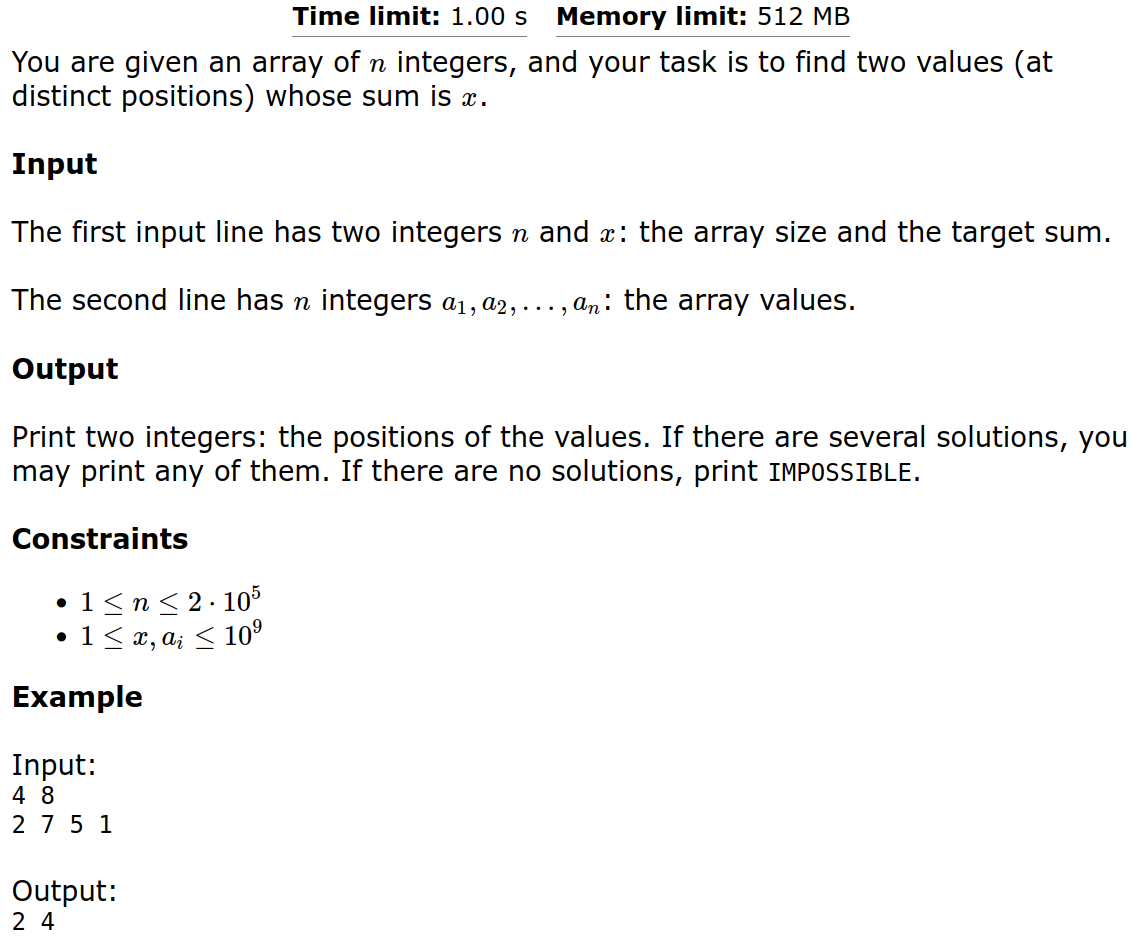}
    \caption{Description of task \textit{CSES 1640 - Sum of Two Values}}
    \label{fig:sumtwo}
  \end{center}
\end{figure}

Figure~\ref{fig:sumtwo} shows the description of the problem \textit{CSES 1640 - Sum of Two Values}~\footnote{\url{https://cses.fi/problemset/task/1640}},
where, given a set of integers, we need to find two of them whose sum equals a specific input value.
This description indicates that we should read the size, given by $n$, and the values of an array, where $n$ may be up to $200,000$. So, a user solution must be fast enough to give the correct answer when $n = 200,000$. Typically, the maximum input values lead an inefficient solution to exceed the time limit, and thus to be rejected.

In the case of problem CSES 1640, a user solution whose running time is $O(n^2)$, i.e., its running time is proportional to the square of the array size, will probably be rejected.

There are several valid solutions for this problem, such as:
\begin{itemize}
    \item Usage of a map to look for the complement of a number;
    \item Sort the values and then use the two-pointer technique to check if we can find the desired sum~\cite{laaksonen2020guide};
    \item Sort the values and then perform a binary search looking for the complement of a number.
\end{itemize}

Thus, CSES contains a comprehensive set of different solutions, using different algorithmic approaches, for a given problem. The site works with several programming languages, but our study focuses on C++, which is the most used language in CSES and in programming competitions in general, \added{and in Java, which is heavily used for enterprise-level development}. So we will investigate, for a set of C++ \added{and Java} solutions, the relationship between execution time and memory usage with energy consumption. 

Besides using several \replaced{algorithmic}{algorithm} strategies, there are also CSES solutions that use strategies to read the input faster. To achieve this, some C++ programs use \texttt{scanf} instead of \texttt{cin} to read the input. Others use \texttt{cin}, but disable the synchronization between the C and C++ standard streams, and the automatic output flushing before a \texttt{cin}. There are also programs that use only the basic \texttt{getchar} function and on top of it define a more high-level reading function. Such reading strategies may be effective when it is necessary to read $100,000$ values or more, as in problem CSES 1640, but should have almost no effect when the input has at most $1,000$ values.

In general, the usage of C++ features, such as templates and the Standard Template Library (STL), varies greatly between CSES programs. There are programs that make heavy use of them, while others do not, and are essentially a C program. Moreover, the C++ solutions from CSES seldom define a new class.

After this overview of programming competition sites focused on CSES, in the next section we discuss our approach to selecting solutions from CSES and measuring its energy consumption.
\section{Methodology}
\label{sec:method}

\subsection{Data collection} 
\label{sec:datacolection}

The CSES problem set has 300 exercises, from which we selected 15. During this process, we tried to select problems that would demand a high computational
effort, so the associated \added{C++ and Java} solutions would run for a reasonable amount of time.

In practice, considering only the largest test cases, we \added{first} sought problems for which the fastest solution ran for at least 0.01 seconds on the CSES server. In a complementary way, the worst solution should have spent nearly one second running on the CSES server. We also tried to select problems involving different topics (e.g., sorting, dynamic programming, mathematics).

For each problem, we selected 30 C++ solutions, where we selected 10 solutions from each one of the following 3 categories:
\begin{itemize}
    \item \textbf{slow:} the top 10 slowest C++ solutions;
    \item \textbf{fast:} the top 10 fastest C++ solutions;
    \item \textbf{rand:} ten C++ solutions chosen at random.
\end{itemize}

\added{These solutions comprise our \textbf{SFR C++} dataset.}

\added{At the time of the data acquisition, the CSES website did not provide a convenient interface to select the fastest solutions in Java. So, to also study the energy profile of Java solutions, we crafted the following new datasets of Java and C++ solutions for each one of the 15 selected CSES problems:}

\begin{itemize}
    \item \added{\textbf{Rand30 C++:} 30 randomly selected C++ solutions (this dataset does not have solutions from the SFR C++ dataset);}
    \item \added{\textbf{Rand30 Java:} 30 randomly selected C++ solutions.}
\end{itemize}

Table~\ref{tab:selectedproblems} shows the CSES ID of each selected problem, \added{its category in the CSES site,} and its associated input/output size. As we can see, the input size varies from problem to problem. There are problems where it is necessary to read hundreds of thousands of values, while in others the input is limited to a few values. Usually, a solution should output a single value, with the notable exception of problem 1071, where a solution should output 100,000 values.

\begin{table}[t]
\centering
\begin{tabular}{llrr}
\toprule
Problem ID    & \added{Category} &  \# Input Values & \# Output Values \\
\midrule
\textbf{1071} &  \added{Introductory Problems}  & 200,000     &   100,000     \\
\textbf{1082} &  \added{Mathematics}            &      1      &         1     \\
\textbf{1084} &  \added{Sorting and Searching}  &  400,000      &         1     \\
\textbf{1140} &  \added{Dynamic Programming}    & 600,000      &         1     \\
\textbf{1158} &  \added{Dynamic Programming}    &   2,000      &         1     \\
\textbf{1621} &  \added{Sorting and Searching}  &  200,000      &         2    \\
\textbf{1632} &  \added{Sorting and Searching}  &  400,000      &         1     \\
\textbf{1634} &  \added{Dynamic Programming}    &     100      &         1     \\
\textbf{1635} &  \added{Dynamic Programming}    &     100      &         1     \\
\textbf{1636} &  \added{Dynamic Programming}    &     100      &         1     \\
\textbf{1639} &  \added{Dynamic Programming}    &  10,000      &         1     \\
\textbf{1640} &  \added{Sorting and Searching}  & 200,000      &         2     \\
\textbf{1642} &  \added{Sorting and Searching}  &   1,000      &         4     \\
\textbf{1643} &  \added{Sorting and Searching}  & 200,000      &         1     \\
\textbf{2185} &  \added{Mathematics}            &      20      &         1     \\
\bottomrule
\end{tabular}
\caption{Selected CSES Problems and the Size of Their Corresponding Test Cases.}
\label{tab:selectedproblems}
\end{table}

For each of the 15 CSES problems, we selected the inputs that would demand more running time from a solution. For some problems, we selected 11 inputs, while for two problems we selected just 2 inputs. On average, we selected around 6 inputs for each problem.

CSES accepts solutions in C++11, C++17 and C++20. As the support for C++20 is still limited and there are not many C++20 solutions in CSES, we only collected solutions for C++11 and C++17. We compiled all C++ collected solutions with \texttt{g++}
and the compiler flag \texttt{std=c++17}.

A C++ solution in CSES is compiled with the optimization flag \texttt{-O2}. As we also wanted to investigate how the GNU C++ compiler optimization flags would affect the energy efficiency of the solutions, we initially compiled each solution with different optimization flags, namely \texttt{-O0} (no optimization), \texttt{-O2} (the ``default") optimization flag and \texttt{-Os} (which enables all \texttt{-O2} optimizations, minus those that frequently increase code size).

In our previous work~\cite{energy2023sblp}, we measured the effect of flags \texttt{-Os}
and \texttt{-O0} for all problems and their solutions. As we could not notice a significantly
different behavior of the energy consumption of the solutions in comparison with the
solutions compiled with \texttt{-O2}, in this extended version we decided to focus only
on the measurements with the flag \texttt{-O2}. Although, for some specific discussions,
such as when presenting Figures~\ref{fig:mmq} and~\ref{fig:outliers}, we measured again
the effect of using the flags \texttt{-Os} and \texttt{-O0}.

\added{In case of Java, the current Java version supported by CSES is 11.0.21.
We did not impose any extra restrictions when randomly picking Java solutions from CSES.}

\subsection{Energy Measurements}
\label{sec:energypmeasurements}

In order to compute the energy consumption of the C++\added{/Java} solutions we used \added{two different measurement approaches, both based on} 
the Running Average Power Limit (RAPL) interface provided by Intel~\cite{weaver2012measuring}. 
RAPL gives consumption values for all cores (energy of all CPU cores) and package, which measures the energy consumption of the entire socket. It includes the consumption of all the cores, integrated graphics, and also the uncore components (last level caches, memory controller) \cite{khan_rapl_2018}. In our work, we considered \added{only} the package value.

\added{Our first measurement approach, which we will call \textbf{\systime},} it
was used by previous works~\cite{pereira2017energy,pereira2021ranking}, which investigated the energy consumption among different programming languages\footnote{The original measuring framework is publicly available at~\url{https://sites.google.com/view/energy-efficiency-languages}}.

\added{Our second measurement approach is based on the well-known Linux tool \textbf{\perf}~\cite{perf}, which provides several hardware and software metrics related to the execution of a program.}

\replaced{Our measuring framework executes a given solution for all the corresponding test cases we have selected, and can perform the measurements using either \systime\, or \perf.}
{As each CSES solution should be executed for several inputs, we adapted
the measuring framework to execute a given solution for all the corresponding
test cases we have selected.}

We measured the energy consumption and execution time of each solution (with each optimization flag) for a certain problem by executing it, for each corresponding selected input, 10 times. Then we dropped the \replaced{execution that consumed the largest energy amount, as also the one that consumed the least energy amount}{slowest and fastest executions}, resulting in 8  measurements. They were combined by computing mean execution time, $t_i$, and mean energy consumption, $c_i$, (and their variances $\sigma_{t_i}$ and $\sigma_{c_i}$) for the $i^{th}$ solution. 

The \deleted{same} measurements were performed in \replaced{three}{two} different desktop computers with Intel processors,
which we labeled as \textbf{HPELITE}, \textbf{HPTHINK} \added{and \textbf{XEON}}. In Table~\ref{tab:comp} we describe the configuration of each computer. In comparison with our previous work~\cite{energy2023sblp}, we updated the operating system and C++ compiler from machine HPELITE. The previous operating system was Ubuntu 18.04.4 LTS (now is Ubuntu 22.04.4 LTS), while the previous version of \texttt{g++} was 7.5.0 (now is 11.4.0). The configuration of machine HPTHINK remains the same\added{, with a minor update in the Ubuntu version}. \added{Moreover, we introduced a new machine, XEON, we are also measuring Java programs, and we are also using \perf\, for some measurements.}

\begin{table}[t]
\centering
\begin{tabular}{lrrr}
\toprule
Config.    &  HPELITE              &      HPTHINK           &       \added{XEON}    \\
\midrule
CPU        &   i5-7500             &   i5-2400             &  \added{Intel-Xeon W-2102}   \\
CPU Freq.  &   3.4 GHz             &   3.1 GHz             &  \added{2.9 GHz}            \\
\added{Cores}   &      \added{4}  &    \added{4}           &  \added{4}  \\
RAM        &   8G DDR4             &   4G DDR3             &  \added{16G DDR4}   \\
RAM Freq.  &   2400 MHz           &    1333 MHz            &  \added{2666 MHz}   \\
L3 cache   &   6 MB                &   6 MB                 &  \added{8 MB} \\
OS         &   Ubuntu 22.04.4 LTS  &   Ubuntu 22.04.4 LTS   &  \added{Ubuntu 22.04.4 LTS} \\
g++        &   11.4.0              &   11.4.0              &   \added{11.4.0} \\
\added{OpenJDK}   & \added{11.0.24}  & \added{19.0.2}   &  \added{11.0.24} \\
\added{perf}      & \added{6.5.13}   & \added{6.8.12}   &  \added{6.5.13} \\
\bottomrule
\end{tabular}
\caption{Configuration of Machines \textbf{HPELITE}, \textbf{HPTHINK} and \textbf{XEON}}
\label{tab:comp}
\end{table}

\added{Machine XEON is a desktop server, so we can see in Table~\ref{tab:comp}
that its configuration is superior to the configuration of machines HPELITE and HPTHINK,
which are desktop clients.
Still} according to Table~\ref{tab:comp}, we can see that HPELITE has a faster processor and also a faster RAM memory than HPTHINK.

When measuring the running time and the energy consumption of programs
we used only the terminal mode. We always restart the system when beginning a
measurement. In our previous work, after a restart, we could perform measurements
for several problems. In the current work, after a restart, we only
measure the C++\added{/Java} solutions of a single problem, and using a given
optimization flag \added{in case of C++}. Thus, now all measurements were carried out right
after restarting the operating system.

\added{Moreover, in some experiments we also measure the impact of using a single core.
Although the C++ and Java solutions we collected are single-threaded, the energy consumption
related to the other cores may have an impact, specially in case of Java programs,
where some tasks, such as the garbage collector, may run in a different core.
To disable a core, we execute the following command:}
{\small
\begin{lstlisting}[numbers=none,caption=\added{Example of Command to Disable a Core.}, label={lst:coreoff}]
sudo echo 0 > /sys/devices/system/cpu/cpu1/online
\end{lstlisting}
}

The \added{\systime} measuring framework \deleted{that we used} executes the solutions via a call to the C \texttt{system} function. The command executed by
\texttt{system} performs as many executions of a solution as there are corresponding selected inputs. For example, considering the executable file \texttt{a.out} associated with a \added{C++} solution, in case there are 3 inputs, the command executed by \texttt{system} is similar to the one shown in Listing~\ref{lst:command}:
{\small
\begin{lstlisting}[caption=Example of \texttt{system} Command Used to Measure Wall-Clock Time., label={lst:command}]
./a.out < input01 > /dev/null 2>&1  && 
./a.out < input02 > /dev/null 2>&1  && 
./a.out < input03 > /dev/null 2>&1
\end{lstlisting}
}

We used the operator \texttt{<} to read the values from the input files, whereas we used\, \texttt{> /dev/null 2>\&1}\, to discard the standard and the error outputs. This discarding should diminish the impact of the output commands on the running time.
The running time and energy consumption associated with the execution of this command consist of a single measurement.

The \deleted{measuring framework} main program \added{of \systime}\, then computes the difference between the wall-clock
time before and after the system call outlined at Listing~\ref{lst:command}.

In this extended version of our previous work~\cite{energy2023sblp},
we updated the measuring framework in order to also measure the
CPU time of C++ solutions\added{, and to measure Java solutions too}.
To measure the CPU time of a solution, we use the
\texttt{time} command provided by the operating 
system~\footnote{To calculate the CPU time, we summed up the user CPU time and the system CPU time reported by the \texttt{time} command.}.
Listing~\ref{lst:commandtime} outlines the system call
performed in such case:
{\small 
\begin{lstlisting}[caption=Example of \texttt{system} Command Used to Measure CPU Time., label={lst:commandtime}]
{time ./a.out < input01 > /dev/null 2>&1;} 2>> tmp &&
{time ./a.out < input02 > /dev/null 2>&1;} 2>> tmp &&
{time ./a.out < input03 > /dev/null 2>&1;} 2>> tmp
\end{lstlisting}
}

As we can see in Listing~\ref{lst:commandtime}, we used the command
\texttt{time} to measure the CPU and wall-clock times of a given C++ solution
against each input \replaced{file}{program}, then we appended the results reported by
\texttt{time} to a temporary file \texttt{tmp}. Later, after running all
the tests, we process the \texttt{tmp} file to get a C++ solution
overall CPU and wall-clock times.

\added{In case of a Java program, we use
\texttt{java MyClassFile} instead of \texttt{./a.out}
to execute a solution.}

\added{The overall approach is similar when we use the \perf\,
measurement framework. The main difference is that we should
execute the \perf\, tool, with the flags to track energy and time events,
once for each test case. Below, Listing~\ref{lst:commandperf}, 
which abstracts some issues related to input/output, shows how
to measure the energy consumption and the execution time associated
with a single test case using the \perf\, tool:}
{\small 
\begin{lstlisting}[numbers=none,caption=\added{Example of Measurement Using \perf.}, label={lst:commandperf}]
perf stat -x ';' -e power/energy-pkg/,user_time,system_time --all-cpus ./a.out < input01
\end{lstlisting}
}

By looking the results, we could see that usually the difference between the
CPU and the wall-clock times reported by \texttt{time} was negligible, less
than 10 milliseconds. However, for some C++ solutions this difference was consistently
higher than 50 milliseconds for all the 10 measurements.
We then performed a manual analysis to understand such differences.

During the analysis, we could see that some C++ solutions were producing segmentation fault errors when running against some inputs.
We identified two causes for these errors:
\begin{enumerate}
    \item \textbf{Insufficient Stack Size:} the default stack size limit
    for our both machines (HPELITE and HPTHINK) was 8192 kb.
    This was not enough for some solutions that were defining a local array
    after reading the number of elements that the array should store.
    As such solutions were accepted by CSES, probably the CSES server
    must have a larger stack size limit.
    We fixed this issue by increasing the stack size through the
    command \texttt{ulimit -s unlimited}.

    \item \textbf{Unsupported CPU Instructions:} some solutions use a pragma such as \texttt{\#pragma GCC target "avx2"} to ask the compiler to use a more specific instruction set for vectors. However, the extended instructions provided by \texttt{avx2} are not supported by machine HPTHINK. We solved
    this issue by commenting out such pragmas, so this extended instruction set was disabled for both machines. \added{The removal of such pragmas did not affect the validity of the solutions, which the online CSES judge still accepted.}
\end{enumerate}

After these changes, for all solutions, the difference
between CPU and wall-clock times remained below 10 milliseconds
in most of the 10 measurements we performed for each C++ solution.

\subsection{Energy Consumption Profiling}
\label{sec:energyprofiling}

Once we had the mean execution time and consumption of all the solutions to a problem, they were plotted, and grouped by optimization flag and machine. For each one of these groups, we fitted the function $\hat{c} = at$, for a predicted consumption $\hat{c}$ given a time $t$, finding the angular coefficient $a$ using ordinary least squares \cite{burden_alise_2016}, minimizing the sum of the squared errors, $SSE = \sum {e_i}^2 $, where the error, or residual, $e_i$ is given by Equation \ref{eq:residual}. We assume the model $\hat{c} = at$, without the linear coefficient $b$, because, for an execution time of zero, the predicted consumption should also be zero. \added{We considered the possibility that the linear coefficient $b$ might represent the baseline energy of an idle machine, however, this value varied greatly for each problem in the same machine, assuming positive as well as negative values. Hence we decided to use a program with a sleep command, with various sleep durations, for this purpose. With this data, we found the idle consumption slope for each machine, which was then removed from the energy measurements before using the least squares method as mentioned above.}

Since each point belonging to a certain group has variances associated with it ($\sigma_{t_i}$ and $\sigma_{c_i}$ as mentioned before) we also decided to use a weighted least squares method \cite{strutz_data_2015}, minimizing the sum of the weighted residuals, that is $\sum w_i \cdot {e_i}^2 $, where the weight $w_i$ of each point was the inverse of its consumption variance, $w_i = \frac{1}{\sigma_{c_i}}$, or the inverse of the product of its consumption and time variances, $w_i = \frac{1}{\sigma_{t_i} \cdot \sigma_{c_i}}$. However, the results obtained by the weighted least square were not good, varying the computed angular coefficient, $a$, of a given problem for each one of the optimization flags much more than the results given by the ordinary least square. In this scenario, we also had problems defining the outlier identification procedure (which will be discussed further below) that would give consistent results. For these reasons, we will not include in this work the results given by the weighted least squares method.

To quantify how well the acquired data respects a monotonic relationship between execution time and energy consumed we computed, for each group, the Spearman correlation factor \cite{myers_research_2003}, as given by $\rho_S$ Equation \ref{eq:spearman}, where $R(t)$ is a ranking of execution time, $R(C)$ a ranking of consumption, and $\sigma_{R(t)}$ and $\sigma_{R(c)}$ are the standard deviation of the rank variables. It is not necessary to assume that the data follow a normal distribution to use the ordinary least squares method~\cite{williams2013assumptions} or the Spearman factor.

\begin{equation}
    \rho_S = \frac{cov(R(t),R(c))}{\sigma_{R(t)}\sigma_{R(c)}}
    \label{eq:spearman}
\end{equation}



To identify outliers, that is, interesting cases where the relationship of the energy consumption did not behave linearly with execution time, we performed the following: first, we computed the residuals, $e_i$ (eq. \ref{eq:residual}) of the best-fit line $\hat{c} = at$ given by the ordinary least squares method for each point $(t_i,c_i)$ in a given group, where $t_i$ represents execution time and $c_i$ energy consumption for the $i^{th}$ solution analyzed.

\begin{equation}
    e_i = \hat{c_i} - c_i = at_i - c_i
    \label{eq:residual}
\end{equation}
Then we can compute the standard deviation of the residuals, $\sigma_e$ by
\begin{equation}
    \sigma_e = \sqrt{\frac{SSE}{n-2}},
    \label{eq:sde}
\end{equation}
where the sum of the squared errors $SSE = \sum e_i^2$ and $n$ represents the number of points of the data set (in our case, 30 points, since we have 30 solutions for a given problem). Usually, we can consider an outlier any point that has a residual $e_i$ that is greater, in absolute value, than twice the standard deviation of the residuals, that is, $|e_i|>2\sigma_e$ \cite{noauthor_book_2015}. However, in our case, each point $c_i$ is obtained from an average operation, with its given standard deviation $\sigma_{c_i}$, which means it has an associated uncertainty (we ignored the variance in the computed execution time $t_i$). To take this into account, we considered 3 cases of growing confidence that the point is indeed an outlier:
\begin{itemize}
    \item $|e_i|>2\sigma_e$: low confidence outlier;
    \item $|e_i|>2\sigma_e + \sigma_{c_i}$: medium confidence outlier;
    \item $|e_i|>2\sigma_e + 2\sigma_{c_i}$: high confidence outlier.
\end{itemize}

\begin{figure}[ht]
    \centering
    \includegraphics[width=0.60\textwidth]{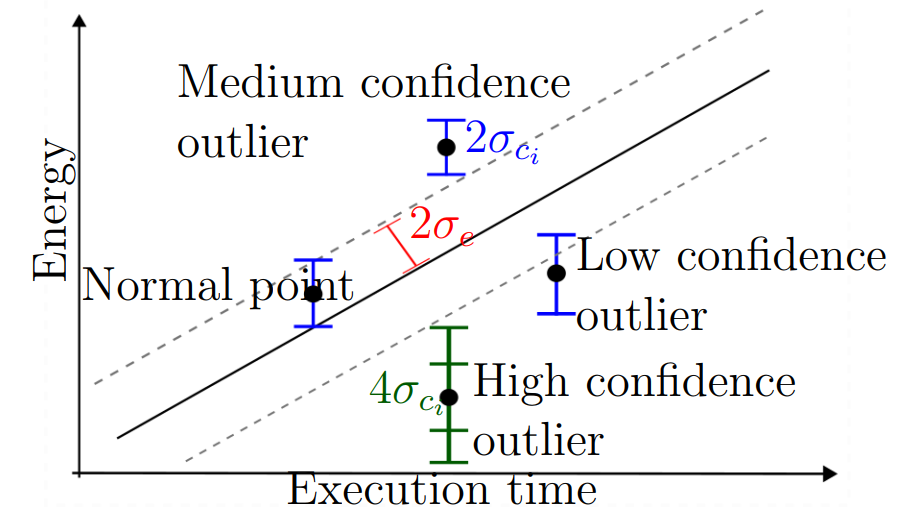}
    \caption{Outlier detection levels for a group of solutions.}
    \label{fig:out}
\end{figure}

Figure \ref{fig:out} shows graphically how to classify data points based on their distance from the best-fit line and each point's uncertainty. If the mean consumption of a point is inside the $2\sigma_e$ area, it is not considered an outlier. If the mean consumption of a point is outside the $2\sigma_e$ area, it is considered at least a low confidence outlier. If the uncertainty of the point, consisting of its value  $\pm \sigma_{c_i}$ (totaling the $2\sigma_{c_i}$ region shown in Figure \ref{fig:out}), is completely outside the $2\sigma_e$ area, it is considered at least a medium outlier. And, if the value of a point $\pm 2\sigma_{c_i}$ ($4\sigma_{c_i}$ region in Figure \ref{fig:out}) is completely out, it is labeled a high confidence outlier. 

With that methodology to classify outliers, we further investigated solutions that were classified, for the same compilation flag, as outliers on both machines where we ran the tests, given that the solution was considered a medium or high confidence outlier in at least one machine.
\footnote{\added{We used this criteria because low confidence outliers are a bit transient
between different executions of our experiment.}}. In those cases, where a solution was an outlier in both machines, we also performed a memory analysis about the heap usage by using the \texttt{valgrind} tool. Since data about memory usage given by \texttt{valgrind} were consistent among repeated executions, including at different machines, we used, for a given solution, the result of a single execution for each selected input, taking note of allocs, frees, total memory used, and memory leaks. 

\subsection{Problem classification}
\label{sec:classification}

Once we computed the angular coefficient $a$ (energy consumption profile) for each one of the problems, as mentioned in Section \ref{sec:energyprofiling}, we conducted a preliminary investigation of the classification potential of this parameter. We wanted to see how good a discriminator this energy consumption profile can be for identifying problems given a set of solutions that solves one of them. 

\added{In this investigation, we used our two C++ data sets (SFR C++ and Rand30 C++),
and} we considered the angular coefficient computed separately in each one of the machines used in the experiment, as well as using only the \texttt{-O2} (the ``default" optimization flag).
Recall that \replaced{these}{for C++ we had} two sets \replaced{have 30 solutions each}{of 30 solutions each: 30 random solutions and a set divided into 3 subsets: slow, fast, and rand}. We trained the classification model for each of these sets\replaced{, which}{. This} gave us two experimental scenarios.


As a test set, we randomly selected a set of 10 solutions for each problem (totaling 15 sets of 10 solutions each). For each one of these sets belonging to the test set, we computed the angular coefficient $a$. Since the number of solutions used in the test set is small, there is a small chance of selecting problems with high running time, meaning most of them will have a small running time (those are more numerous on the website). This might affect the computed slope, as the slowest (and fastest) solutions usually deviate more from the expected consumption profile when compared to solutions with about average execution time. Hence, it is expected to have better results when training with the random set of solutions.

Notice that there is no guarantee that the computed slope $a$ is unique for each problem --- slope $a_i$ given by a set of solutions of the $i^{th}$ problem is not necessarily distinct from the slope $a_j$ of the $j^{th}$ problem. However, the consumption profile might limit the search space, indicating, for example, that the test problem has a high consumption profile, and thus you should consider only the problems in the training set that also have high slopes. For this reason, after computing the parameter $a$ of a given set belonging to the test set, we compared this value to all the values given by the training set. Considering the $n^{th}$ closest values, and varying $n$, we could see how well we could select a subset of the training problems that contained the correct classification of the test problem.

\section{Results and Discussion}
\label{sec:results}

In this section, we will present and discuss the results obtained after analyzing the data acquired. Our data and results are publicly available in a git repository~\cite{medeiros2024dataset}.

Section \ref{sec:resprofile} presents graphical examples of energy consumption profiles and addresses the question of the linearity relation between energy consumption versus time\replaced{. Then}{, and}, in Section~\ref{subsec:outmem}, we discuss the \added{usage of a linear function that estimates the energy consumption of a program, based on its execution time, to identify programs with abnormal energy consumption and} we talk about the association of \replaced{these programs}{outliers} with memory management. Then, Section \ref{subsec:resclass} shows the results obtained by the proposed classification method based on slopes. Finally, in Section~\ref{subsec:threats}, we comment on some issues that may have influenced our results.

\subsection{Energy Consumption Profiling}
\label{sec:resprofile}

\begin{figure}[htb]
  \begin{center}
    \leavevmode
    \includegraphics[width=0.9\textwidth]{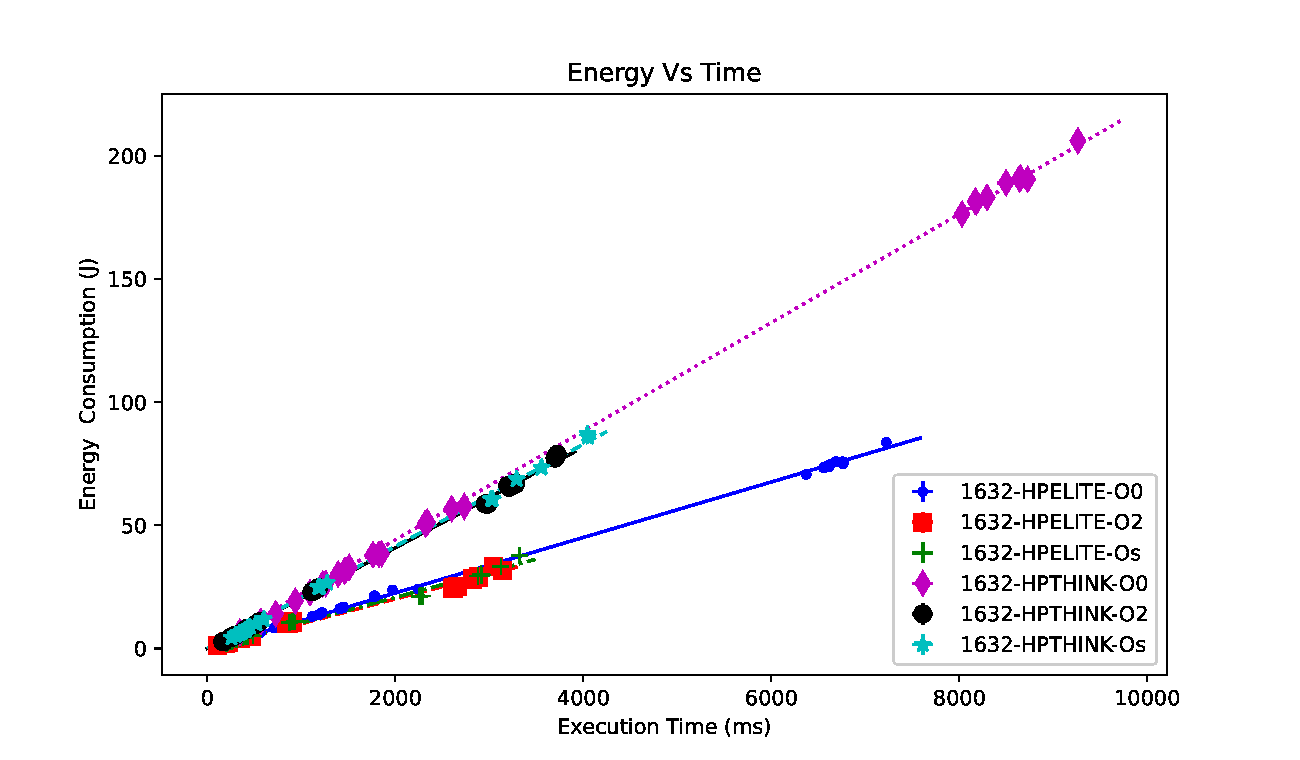}
    \caption{Data points and linear regression of solutions to the problem CSES 1632.}
    \label{fig:mmq}
  \end{center}
\end{figure}

\replaced{In this section, we will discuss the straight line given by the \replaced{least squares}{MMQ} method that relates energy consumption and CPU time. Section~\ref{sec:flags} discusses the impact of compilation flags on this line in case of programs from the SFR C++, and  Section~\ref{sec:evalrand} shows, in the context of the Rand30 C++ and Rand30 Java data sets, how this line changes depending on factors such as the measurement framework and the available CPU cores.}
{This discussion considers the results obtained by using the SFR data set.}

\added{As the following discussion shows, it seems that we can give a positive answer
for our first research question~\ref{rq:question1}:
\textit{By using a small set of C++ and Java programs, aimed to solve a given problem, can we get a linear function that fits the corresponding energy consumption and execution time data points?}}

\subsubsection{\added{Evaluating the Impact of Compilation Flags on the Energy Profile for the SFR C++ Data Set}}
\label{sec:flags}

\added{To evaluate the impact of optimization flags on the energy consumption of C++ programs, we performed an experiment in machines HPELITE and HPTHINK using the \systime\, measurement framework. As mentioned in Section~\ref{sec:method}, we compiled each solution from the SFR C++ data set with different optimization flags: -O0, -O2 and -Os.
}

Figure \ref{fig:mmq} shows, as an example, the correspondence between running time and energy consumed for the set of solutions of the problem CSES 1632. \replaced{The measurements were obtained in two different desktop machines (labeled HPELITE and HPTHINK), and each solution was compiled with three different optimization flags, as discussed in Section \ref{sec:method}, totaling 180 points.}{As we compiled each of the 30 solutions with 3 optimization flags on 2 different machines, Figure~\ref{fig:mmq} represents 180 data points.} We can see that the machine HPELITE is faster when compared to machine HPTHINK, \replaced{and that it also consumes}{thus consuming} less energy when running a given solution. 

\added{More importantly, we can see that in both machines the usage of optimization flags did not have much impact on the computed energy consumption profile - the angular coefficient given by the
ordinary least squares method was roughly the same.}
We can also observe that solutions compiled with -O2 are faster than \added{the ones compiled with} -Os, which in turn \replaced{are}{is} faster than \added{the ones compiled with} -00, as expected. Time is expressed in milliseconds, while energy consumption is in Joules.

As you can see in Figure \ref{fig:mmq}, all the six groups analyzed in this figure were fitted nicely by a linear function, with Spearman correlation coefficients above 0.99. This kind of behavior happened for almost all data collected. 
Considering only the \texttt{-O2} flag, the Spearman correlation coefficients obtained for all the 30 groups were often (in 77\% of the cases) above 0.98, and the lowest value, obtained once, was 0.93.

\deleted{Based on this, we can give a positive answer to our first research question~\ref{rq:question1}:}
\textit{\deleted{Can we confirm that, for this diverse set of mined C++ solutions, there is a strong correlation between energy consumption and execution time?}}

\subsubsection{\added{Evaluating the Energy Profile of Randomly Picked C++ and Java Programs}}
\label{sec:evalrand}

\added{In this Section, we will evaluate the energy profile of C++ and Java solutions picked at random. To such evaluation, we will use the Rand30 C++ data set, compiled only with the -O2 optimization flag, and the Rand30 Java data set.}

\added{First, we will use the \systime\, measuring framework to measure the energy consumption of the solutions of both data sets on machines HPELITE and HPTHINK.
Table~\ref{spearman-rand-systime} shows the Spearman correlation coefficients for the Rand30 C++ and Rand30 Java data sets.
Column \textit{Mult} has the results when all cores are enabled, while column \textit{Sing} shows the results when a single core is enabled.
}

\begin{table}[ht]
    \centering
    \begin{tabular}{|c|c|c|c|c|c|c|c|c|}
            \hline
            ID & \multicolumn{4}{c|}{HPELITE} & \multicolumn{4}{c|}{HPTHINK} \\ \hline
               & \multicolumn{2}{c|}{C++} & \multicolumn{2}{c|}{Java} & \multicolumn{2}{c|}{C++} & \multicolumn{2}{c|}{Java} \\ \hline
     &    Mult     &  Sing      &   Mult     & Sing  &  Mult      &  Sing       & Mult       & Sing\\ \hline
1071 &  0.966	   & 0.956   	& \rd{0.879} & 0.992 & \rd{0.871} & \rd{0.869}	& 0.983	     & 0.989 \\ \hline
1082 &  0.992	   & 0.990	    & 0.991	     & 0.995 & 0.992      & 0.992	    & 0.984	     & 0.990 \\ \hline
1084 &  0.990	   & 0.996	    & \rd{0.788} & 0.996 & \rd{0.946} & 0.959	    & \rd{0.800} & 0.993 \\ \hline
1140 &  0.995	   & 0.988	    & \rd{0.942} & 0.996 & 0.988	  & 0.987	    & \rd{0.801} & 0.980 \\ \hline
1158 &  0.996	   & 0.998	    & 0.983	& 0.992	& 0.992	& 0.996	& 0.970	& 0.995 \\ \hline
1621 &  0.975	   & 0.978	    & 0.968	& 0.994	& 0.974	& 0.969	& 0.988	& 0.995 \\ \hline
1632 &  0.973	   & 0.981	    & \rd{0.668}	& 0.990	& 0.993	& 0.992	& \rd{0.886}	& 0.992 \\ \hline
1634 &  0.982	   & 0.984	    & 0.990	& 0.994	& 0.983	& 0.984	& 0.979	& 0.985 \\ \hline
1635 &  \rd{0.854} & \rd{0.861}	& 0.975	& 0.974	& \rd{0.901}	& \rd{0.911}	& 0.970	& 0.966 \\ \hline
1636 &  0.987	   & 0.989	& 0.981	& 0.995	& 0.986	& 0.987	& 0.980	& 0.982 \\ \hline
1639 &  0.989	   & 0.991	& 0.991	& 0.993	& 0.989	& 0.987	& 0.958	& 0.994 \\ \hline
1640 &  0.967	   & 0.964	& 0.976	& 0.993	& 0.992	& 0.997	& \rd{0.926}	& 0.997 \\ \hline
1642 &  0.994	   & 0.992	& \rd{0.935}	& 0.991	& 0.999	& 0.996	& \rd{0.936}	& 0.988 \\ \hline
1643 &  0.976	   & 0.969	& \rd{0.861}	& 0.983	& \rd{0.876}	& 0.981	& \rd{0.944}	& 0.988 \\ \hline
2185 &  0.967	   & 0.968	& 0.963	& 0.994	& 0.986	& 0.992	& 0.980	& 0.998 \\ \hline
        \end{tabular}
        \caption{\added{Spearman Correlation for the \textbf{Rand30 C++} and the \textbf{Rand30 Java} Data Sets in Machines HPELITE and HPTHINK using the \systime\, Measurement Framework. Column \textit{Mult} indicates that all cores are enabled,
        while column \textit{Sing} means a single core is enabled.
        We highlighted the entries where the Spearman Correlation was below $0.95$}\label{spearman-rand-systime}}
\end{table}

\added{
Table~\ref{spearman-rand-systime} has 120 entries, and overall (for 100 entries)
the Spearman correlation coefficients were above 0.95. In case of the 20 entries
where the Spearman correlation was below 0.95, we can see that 8 of them are
related to C++, while 12 are related to Java. More importantly, we can
see that most entries with a Spearman correlation below 0.95 (17 out of 20)
were obtained with all cores enabled. Thus, keeping a single core enabled
increased the correlation between energy and time in case of our data set
of non-parallel programs. In our measurements with a single
core enabled, in case of Java, we got a Spearman correlation coefficient
above 0.95 for all problems, while in case of C++ this coefficient was
below 0.95 only for problems 1071 and 1635.
}

\added{
We then performed another experiment to calculate the Spearman correlation
coefficients by using the \perf\, measuring framework. In this new experiment,
we also included machine XEON. Such experiment confirmed the trend
that the correlation between time and energy increases when we keep a single
core enabled when running the single-threaded programs from our data sets.
Table~\ref{spearman-rand-perf} reports the results of this new experiment
when a single core is enabled. 
}

\begin{table}[ht]
    \centering
    \begin{tabular}{|c|c|c|c|c|c|c|}
        \hline
        ID & \multicolumn{3}{c|}{C++} & \multicolumn{3}{c|}{Java}  \\ \hline  
           & HPELITE & HPTHINK & XEON & HPELITE & HPTHINK & XEON \\ \hline
1071 & \rd{0.928} &	\rd{0.943} & 0.960 & 0.996 & 0.993 & 0.977 \\ \hline
1082 &     0.985  & 	0.993  & 0.998 & 0.999 & 0.994 & 0.996 \\ \hline
1084 &     0.972  &	    0.954  & 0.943 & 0.998 & 0.996 & \rd{0.943} \\ \hline
1140 &     0.989  &	    0.994  & 0.998 & 0.992 & 0.987 & 0.994 \\ \hline
1158 &     0.992  &     0.983  & 0.993 & 0.991 & 0.996 & 0.993 \\ \hline
1621 &     0.976  &     0.971  & 0.979 & 0.998 & 0.999 & 0.970 \\ \hline
1632 &     0.981  &     0.995  & 0.991 & 0.994 & 0.983 & 0.980 \\ \hline
1634 &     0.984  &     0.980  & 0.980 & 0.993 & 0.986 & 0.992 \\ \hline
1635 & \rd{0.848} & \rd{0.903} & \rd{0.942} & 0.973 & 0.962 & 0.956 \\ \hline
1636 &     0.986  &	    0.985  & 0.984 & 0.994 & 0.981 & 0.980 \\ \hline
1639 &     0.981  & 	0.981  & 0.993 & 0.994 & 0.989 & 0.986 \\ \hline
1640 &     0.972  & 	0.996  & 0.998 & 0.992 & 0.998 & 0.996 \\ \hline
1642 &     0.990  &     0.998  & 0.994 & 0.990 & 0.991 & 0.997 \\ \hline
1643 &     0.951  & 	0.985  & 0.978 & 0.993 & 0.994 & 0.965 \\ \hline
2185 &     0.961  &	    0.984  & 0.987 & 0.996 & 0.997 & 0.980 \\ \hline           
        \end{tabular}
        \caption{\added{Spearman Correlation for the \textbf{Rand30 C++} and the \textbf{Rand30 Java} Data Sets in Machines HPELITE, HPTHINK and XEON Using the \perf\, Measurement Framework, with a Single Core Enabled.
        We highlighted the entries where the Spearman Correlation was below $0.95$}\label{spearman-rand-perf}}
\end{table}

\begin{figure}[htb]
  \begin{center}
    \leavevmode
    \includegraphics[width=0.9\textwidth]{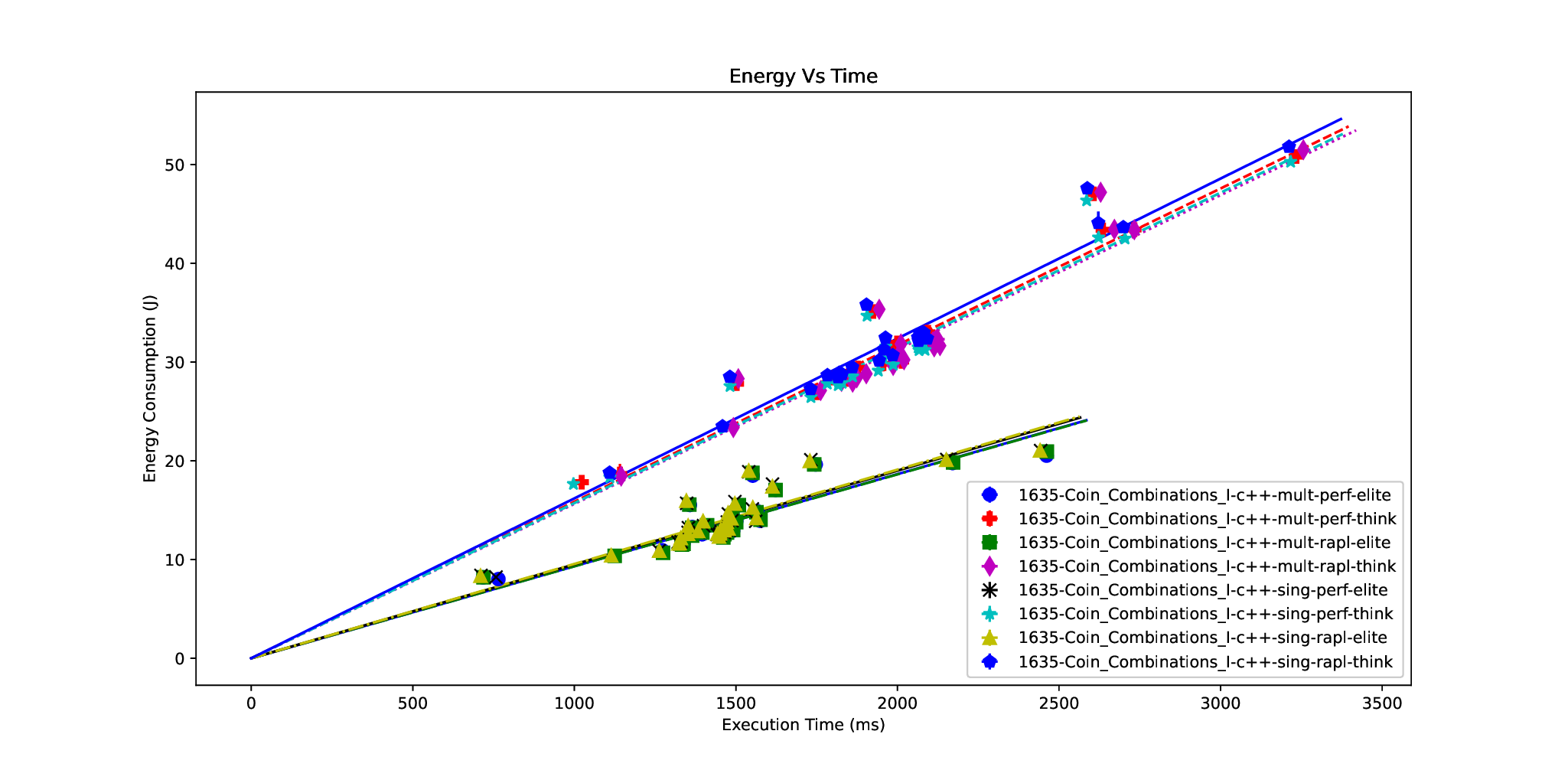}
    \caption{\added{Energy Consumption Profile of Randomly Picked C++ Solutions for Problem CSES 1635. We can see many points with similar execution time and energy consumption.}}
    \label{fig:lowestsperamnelite}
  \end{center}
\end{figure}

\added{
In the case of C++, we can see that the Spearman coefficient for problems
1071 and 1635 was again below 0.95 for machines HPELITE and 
HPTHINK, while in XEON it was below only for 1635.
To better understand this, let us look at the energy profile
of problem 1635, which is shown in Figure~\ref{fig:lowestsperamnelite}.
}

\added{We can see that there are several solutions with similar execution times,
but with a slight difference in energy consumption. 
This is a possible explanation for the low correlation since a small variation in the energy measurement in points with similar execution times can have a strong negative effect on the Spearman correlation coefficient.}

\added{For Java, with single core measurements, the Spearman coefficients were always above 0.97. When all cores are enabled, the Spearman coefficient was above 0.95 most of the time (85\%), however, there were some values as low as 0.67 and 0.78 for HPELITE and HPTHINK respectively.}

\begin{figure}[htb]
  \begin{center}
    \leavevmode
    \includegraphics[width=0.9\textwidth]{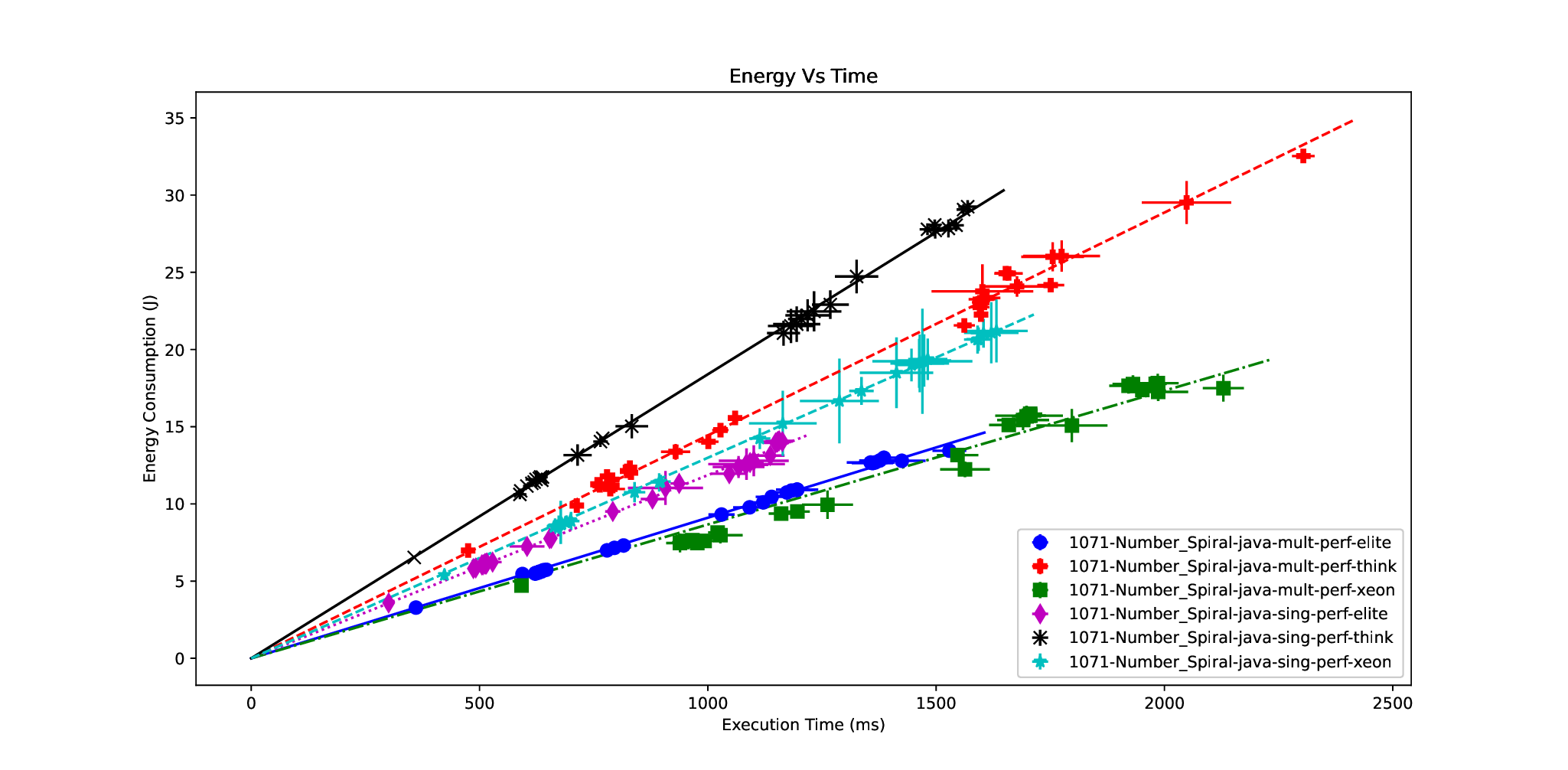}
    \caption{\added{Comparing the Energy Consumption Profile of Java Solutions for Problem CSES 1071 Depending on Core Availability. The measurements using \perf\, indicate that in all machines the usage of multiple cores decreased the slope.}}
    \label{fig:compjava}
  \end{center}
\end{figure}

\added{Figure \ref{fig:compjava} compares the results obtained for problem CSES 1071 with Java using multiple cores and single cores. In this case we can see that the usage of multiple cores decreased the slope for all machines.}

\added{Considering all the problems, in HPTHINK machine, Java mult (using multiple cores) did not increase much the performance when compared to Java single (using a single core). The ratio (slope single)/(slope mult) gave us an average value of 114\%, with a maximum of 130\% for problem CSES 1621 and a minimum of 103\% for problem CSES 1636. For the HPELITE machine the average ratio was 123\%, with a maximum of 159\% for problem CSES 1621 and a minimum of 105\% for problem CSES 1636. In XEON the performance increased 136\% on average, with a maximum of 157\% once again for CSES 1621 and a minimum of 106\% for CSES 1636.}

\added{This shows us that parallelization is machine-dependent as well as problem-dependent. We can see that problem CSES 1621 is highly parallelizable while CSES 1636 is not. }

\begin{figure}[htb]
  \begin{center}
    \leavevmode
    \includegraphics[width=0.9\textwidth]{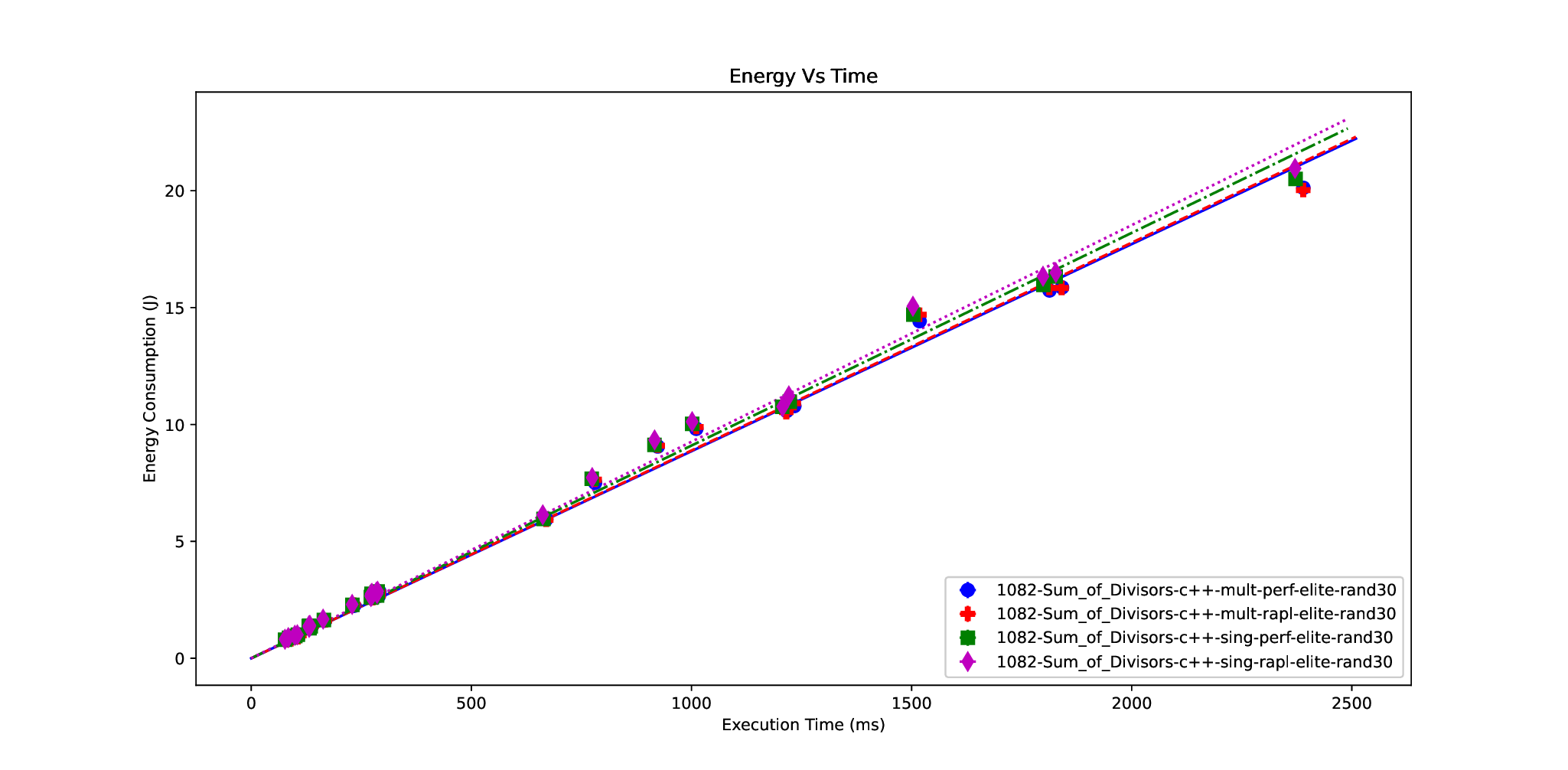}
    \caption{\added{Comparing the Energy Consumption Profile of C++ Solutions for Problem CSES 1082 in Machine HPELITE Using \systime\, and \perf\, Depending on Core Availability. The availability of multiple cores does not have a meaningful impact on the computed slope.}}
    \label{fig:compc}
  \end{center}
\end{figure}

\added{Figure \ref{fig:compc} shows us the results obtained for solutions to problem CSES 1082 with C++ language using multiple and single core and measurements using \systime\, and \perf\ tools. We can see that the results are very similar. Overall, comparing C++ using a single core versus using multiple cores across all the problems shows us that it does not make much difference. The average ratio (slope single)/(slope mult) using \systime\, was 104\%, 102\%, and 97\% for HPTHINK, HPELITE, and XEON, respectively. Regarding the measurements with \systime\, or \perf\, the results as also similar, with values varying about 1\%.}

\added{
To investigate the difference in measurement of execution time using both wall-clock and CPU times, we computed the Spearman correlation coefficient between them. When measuring with  \systime\, for C++, both for single and multi core, 
the Spearman correlation coefficient between the wall-clock and CPU times measured was 
very high, above 0.99 in all measurements (for SFR C++, Rand30 C++ and test sets, in all the machines).
}

\added{
Using \perf\ once again for C++, single and multi core, in all machines and sets, the coefficient
was above 0.95 90\% of the time, and always above 0.80. The mean value of the correlation was
0.98.
}

\added{
The measurements for Java single core, using both  \systime\, and \perf\ for Rand30 Java set, had an Spearman correlation coefficient above 0.99 in all measurements.
}

\added{
The measurements for Java multi core (for Rand30 Java set) using both  \systime\, and \perf\, were the ones with the lowest values for the Spearman correlation coefficient between wall-clock and CPU times. Such coefficient  was above 0.95 in only 50\% of the measurements,
with a mean value of 0.87, with values as low as 0.45 (HPELITE), 0.64 (HPTHINK) and 0.39 (XEON). 
All the lowest values were for problem CSES 1642. Disregarding this problem makes all the Spearman correlation coefficients values above 0.7.
We think this could be explained by the fact that different Java solutions to the same problem may have a similar wall-clock running time, but a different CPU time running time, depending on their usage of the multiple cores. For example, in case of problem 1642, the solution
\texttt{entry\_1761344.java} took about 1400 ms of wall-clock time and 2900 ms of CPU time,
while solution \texttt{entry\_3785822.java} took about the same amount of wall-clock time,
but took around 200 ms of extra CPU time to execute.}

\subsection{\replaced{Identifying Programs with an Abnormal Energy Consumption Profile}{Outliers Analysis}}
\label{subsec:outmem}


%

\begin{figure}[htb]
  \begin{center}
    \leavevmode
    \includegraphics[width=0.8\textwidth]{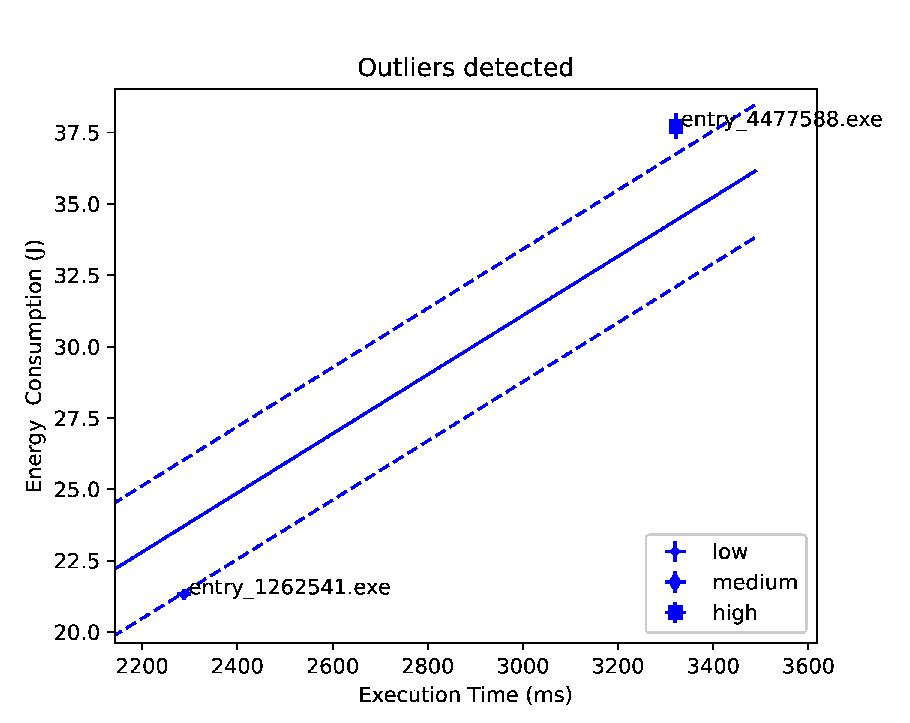}
    \caption{Outliers for solutions to problem CSES 1632 executed in HPELITE with flag -Os. The dashed lines show $\pm 2\sigma_e$ region, and the error bars show $\pm \sigma_c$ region of two detected outliers.}
    \label{fig:outliers}
  \end{center}
\end{figure}


\added{In this section we will discuss if by computing an estimate of the expected energy consumption of a program, given its execution time, can help us to identify programs with an abnormal energy consumption. We will do an analysis considering the \textbf{SFR C++} data set and the machines HPELITE and HPTHINK. Moreover, we will use the \systime\, measuring framework.}

\added{
In Section~\ref{sec:flags}, we saw that the experiments performed on these machines considering such data sets showed a strong correlation between execution time and energy consumption, so the data points that represent the energy consumption of a program were fitted well by the straight line computed by the least square method.
}

Although the majority of points collected showed very good linearity, we were able to find some outliers. In the example of Figure \ref{fig:mmq}, for problem CSES 1632, there were a total of 7 outliers detected (4\%  of the total). Two detected outliers for the group of solutions executed in HPELITE machine with optimization flag -Os are shown in Figure \ref{fig:outliers}, for example. One of them is a low confidence outlier located just below the tolerance zone (which means that it consumed less energy than expected for a code with its execution time). A high confidence outlier is located above the tolerance zone, meaning this solution consumed more energy than would be expected.

It is worth noting that the slopes of the fitted lines of the solutions executed in the faster machine, HPELITE, are smaller than the slopes of the lines that fitted the solutions that ran in the slower machine, HPTHINK. 
This means that not only programs took longer to execute in the slower machine (as shown by a lot of points in the upper right of Figure \ref{fig:mmq}), as expected, but also that an increase in execution time accounts for around twice as much energy consumed when compared to a similar increase in time in HPELITE. However, we can not conclude that slower machines have steeper slopes. This will be defined solely by how efficient (power needed to run the whole system) a machine is. 

Overall, when considering all the problems and all their respective solutions with \texttt{-O2} (15 problems $\times$ 30 solutions $\times$ 2 different machines = 900 data points), a total of 50 (5.6\%) points were tagged as outliers. Of those, 38\% were labeled as low confidence, 14\% as medium, and 48\% as high ones. From these 50 outliers, 28 are related to measurements performed at HPELITE
while 22 of them are related to measurements carried out at HPTHINK.

As explained in Section \ref{sec:method}, we selected for further analyses the solutions compiled with the same compilation flag (\texttt{-O2}) that were tagged as outliers on both machines and in at least one machine the solution was tagged as a medium or high confidence outlier. This led us to the analysis of 8 solutions (16 data points) which were outliers on both machines. 


The distribution of these 8 outliers among the 15 CSES problems was as follows,
where the outliers above, respectively below, the line in one machine were also always above, respectively below, in the other one:
\begin{itemize}
    \item \textbf{0 outlier:} CSES 1071, CSES 1140, CSES 1634, CSES 1636, CSES 1639, CSES 1642, CSES 2185
    \item \textbf{1 outlier:} CSES 1082, CSES 1084, CSES 1158, CSES 1621, CSES 1632, CSES 1635,
                              CSES 1640, CSES 1643
\end{itemize}
As we can see, \replaced{most of CSES problems also had only one outlier}{for most CSES problems we had only one outlier}. 
In our previous work~\cite{energy2023sblp}, most of CSES \added{problems analyzed} also had only outlier, although
the amount of outliers selected for analysis according to the same criteria was higher (13 outliers).
From our 8 current outliers, 7 of them were also classified as outliers in our
previous work.
Probably this reduction in the amount of outliers was due to resolving the issues related to insufficient
stack size and unsupported CPU instructions (see Section~\ref{sec:energypmeasurements}),
which might be affecting the proper computation of the energy consumption profile
of some problems. 

For each outlier, we used \texttt{valgrind} to check if its heap memory usage was different from the solutions with a similar running time. In this case, we could try to associate the different energy consumption with different memory usage. \deleted{This will be discussed in the next section.}

Overall, when we compare the dynamic memory usage of the outliers with the solutions with a similar running time, we could consider that there is a meaningful difference for all the 8 outliers.
Sometimes this different memory usage \added{(e.g., number of alloc/deallocs, amount of allocated memory)} leads to an energy consumption higher than what was expected, and others lead to lower energy consumption.

In our previous work~\cite{energy2023sblp}, we had 2 out of 13 outliers whose 
abnormal energy consumption did not seem related to memory management. 
However, they were not identified as outliers in our current work.
In fact, one of them was still tagged as an outlier, but only in
a single machine.

In the case of problem CSES 1640, whose description we presented in Figure~\ref{fig:sumtwo}, the outlier program was consuming more energy than expected. The usage of \texttt{valgrind} revealed that it was allocating much more memory (between $17\times$ and $100\times$ more) than the solutions with a similar running time, and doing this by performing a higher number of operations to allocate memory (between $2,000\times$ and $1,263,200\times$ more).

Thus, probably this unexpectedly high energy consumption is due to heavy memory usage and how it is allocated. In this case, after inspecting the source code we could find an issue related to the \texttt{comp} function shown below in Listing~\ref{lst:cmpcopy}:
\begin{lstlisting}[language=C++, caption=Function \texttt{comp} Makes a Copy of its Parameters Whenever it is Called by Function \texttt{sort}., label={lst:cmpcopy}]
bool comp(vector<int> a, vector<int> b) {
    return a[0] < b[0];
}
\end{lstlisting}

In C++, by default, a \texttt{vector} passed as an argument to a function is copied. Thus, each time we call function \texttt{comp} we need to copy two vectors. As function \texttt{comp} is called by the \texttt{sort} function, which needs to sort a vector with 200,000 elements, this results in more than 6,000,000 heap allocations. This huge amount of allocations may become a performance bottleneck~\cite{pinto2017forkjoin,pereira2021ranking}.

A quick fix to improve the memory usage, which will also improve the running time and the energy consumption, is to change \texttt{comp} to receive its parameters by reference, as follows: 
\begin{lstlisting}[language=C++, caption=Changing Function \texttt{comp} to Receive its Parameters by Reference., label={lst:cmpref}]
bool comp(vector<int> &a,
          vector<int> &b) {
    return a[0] < b[0];
}
\end{lstlisting}

After this change, the heap memory usage dropped by more than 80\% and the number of allocs dropped by more than 90\%.
We repeated the measurements taking into account this modification, and we could see that the improved solution decreased
its time and energy consumption to around 1/3 of the original solution. Moreover, the new solution is not reported as an outlier for any of the machines.

\subsection{Classification Based on Slope}
\label{subsec:resclass}

\added{In this section, we discuss a preliminary approach to classify a program
based on its energy profile. 
The discussion below considers the \textbf{SFR C++} and the \textbf{Rand30 C++} data sets, and the machines HPELITE and HPTHINK, as well as the \systime\, measuring framework.}

\added{
Usually, the algorithm classification problem consists
on finding which algorithm a program implements among a given finite set of algorithms.
A good classification algorithm is essential in applications such as 
malware identification~\cite{malware2013}. Unlike most approaches~\cite{classification2023cgo},
ours does not depend on a program source code, relying only on the energy consumption profile
produced by an executable program. 
}

As mentioned in Section~\ref{sec:classification}, the energy consumption profile of a set of solutions to a given problem might not be unique. If we take, for example, the Rand30 C++ training set for the HPELITE machine, the computed slopes varied from 0.00909 (1082 - Sum of Divisors) to 0.01253 (1643 - Maximum Subarray Sum). By normalizing the slopes we get a minimum slope of 1.000 and a maximum one of 1.379. The average difference between the normalized slopes of the 15 different problems was 0.027. The two closest problems had slopes 1.252 (1071 - Number Spiral) and 1.255 (1636 - Coin Combinations II), making them hard to tell apart (in this specific machine). If the training set consisted of a lot more than 15 problems, the mean distance between slopes could decrease, making it even harder to distinguish problems based on the slope. \added{This might happen because with a greater population of problems, and considering that the minimum and maximum slopes will probably not change much, there will be more problems distributed in between those values, becoming harder to tell them apart.} However, we wanted to investigate how well we could use the slope to classify problems or at least reduce the search space.

\begin{figure}[!htb]
  \begin{center}
    \leavevmode
    \includegraphics[width=0.80\textwidth]{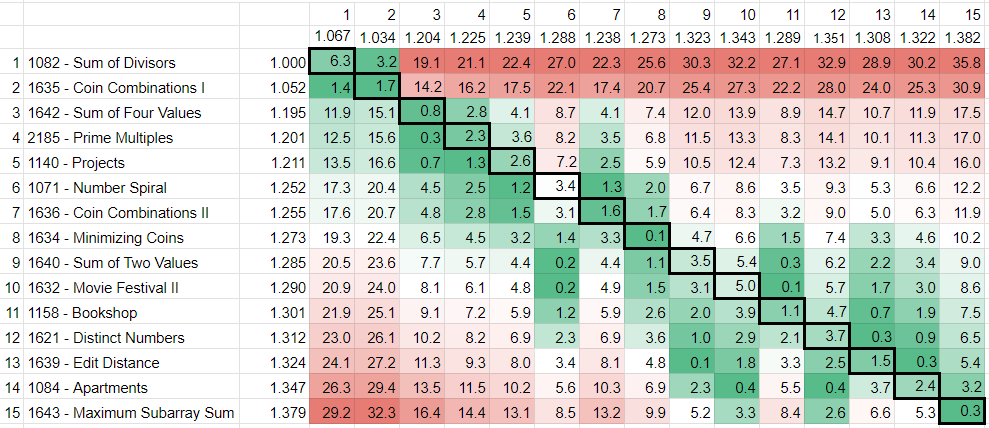}
    \\(a)\\
    \includegraphics[width=0.80\textwidth]{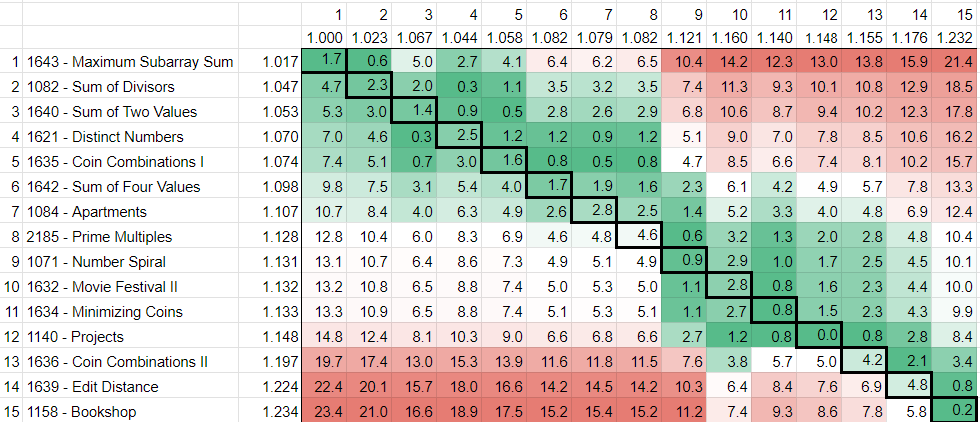}
    \\(b)
    \caption{Classification table of problems using Rand30 C++ set for training in a) HPELITE and b) HPTHINK machine. Cells show the relative difference between computed normalized slopes. If the smallest value in each column is in the diagonal, it means a perfect classification. Green cells represent smaller values and red bigger values.}
    \label{fig:thinkrand30}
  \end{center}
\end{figure}

Figure \ref{fig:thinkrand30} shows the comparison (relative difference in percentage) of the normalized slopes obtained by the 15 test sets (columns) and the normalized slopes obtained by the training sets (Rand30 C++) (lines) in both machines. If the smallest value is in the main diagonal, it means we perfectly identified the test problem. If the smallest values are concentrated around the main diagonal, that means we can reduce the search space\added{, which can be useful to improve the bounding phase of a branch-and-bound classification algorithm~\cite{bb-clustering,bb-survey}}. We can see that the smallest values are mostly concentrated around the main diagonal, showing that the classification method based on the energy consumption profile can indeed distinguish between problems to a certain degree.   

Table \ref{tab:thinkrand30} shows the classification success by selecting the $n^{th}$ smallest values in Figure \ref{fig:thinkrand30} for each column. For a given $n$, we consider it a success if the correct problem belongs to the selected set for each column.  For example, for $n=4$, Rand30 C++ has 11 successes for both machines. That means that if we select four problems with the nearest slopes to a given test problem, the correct one will be one of these four problems 11 times out of 15. We can see that the random training - Rand30 C++ - achieved better results. With this training set, both machines achieved 100\% success rate for $n=7$, which means we can reduce the search space by about half. For $n=1$ the success rate was 20\% and 27\% for HPELITE and HPTHINK machines, respectively, when compared to 7\% (one out of 15) of guessing the right problem by chance. 

\begin{table}[t]
\centering
\begin{tabular}{l|cc|cc}
\toprule
 & \multicolumn{2}{c|}{HPELITE} & \multicolumn{2}{c}{HPTHINK} \\
$n$  & SFR C++ &  Rand30 C++ & SFR C++ &  Rand30 C++  \\
\midrule
1 &  1  &  3  &  0 & 4  \\
2 &  3  &  6   &  4 & 6  \\
3 &  5  &  9   &  4 & 10  \\
4 &  6  &  11  &  5 & 11  \\
5 &  7  &  11  &  8 & 12  \\
6 &  8 &  14  &  10 & 14  \\
7 &  8 &  15  &  11 & 15  \\
8 &  10 &  15  &  11 & 15  \\
9 &  11 &  15  &  12 & 15   \\
\bottomrule
\end{tabular}
\caption{Number of success, for different machines and different training sets when varying the number $n$ of selected nearest problems to the test problem slope.}
\label{tab:thinkrand30}
\end{table}

\begin{figure}[!htb]
  \begin{center}
    \leavevmode
    \includegraphics[width=0.85\textwidth]{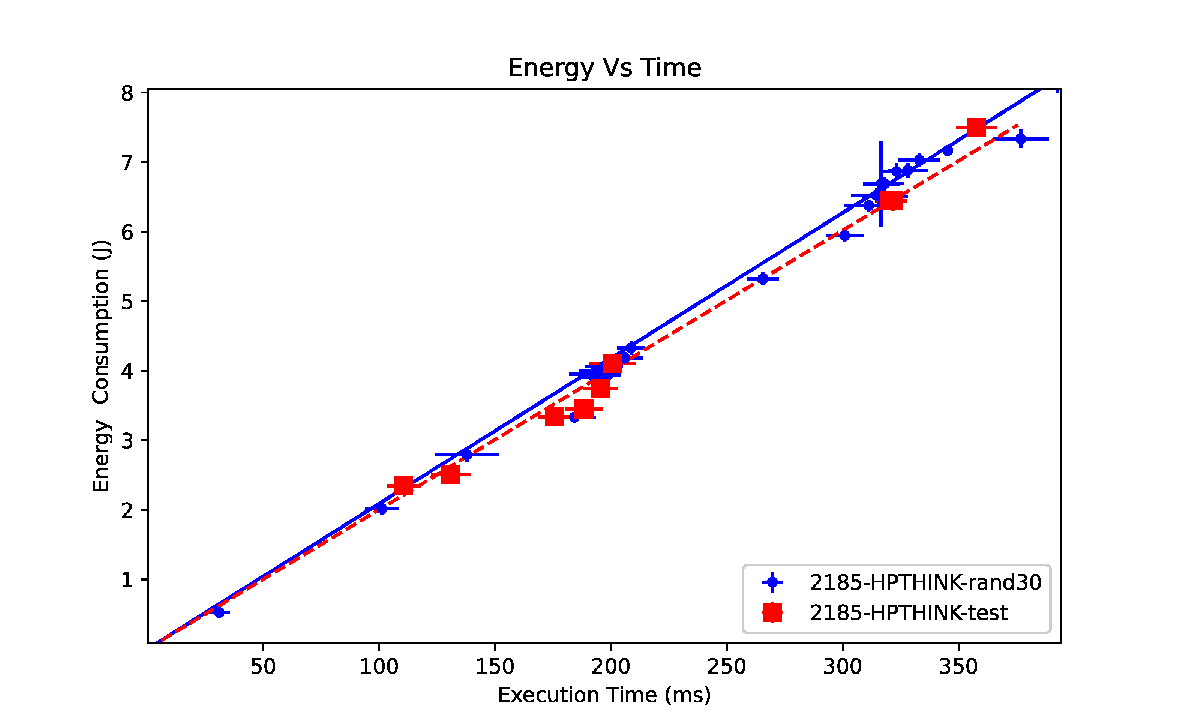}
    \\(a)\\
    \includegraphics[width=0.85\textwidth]{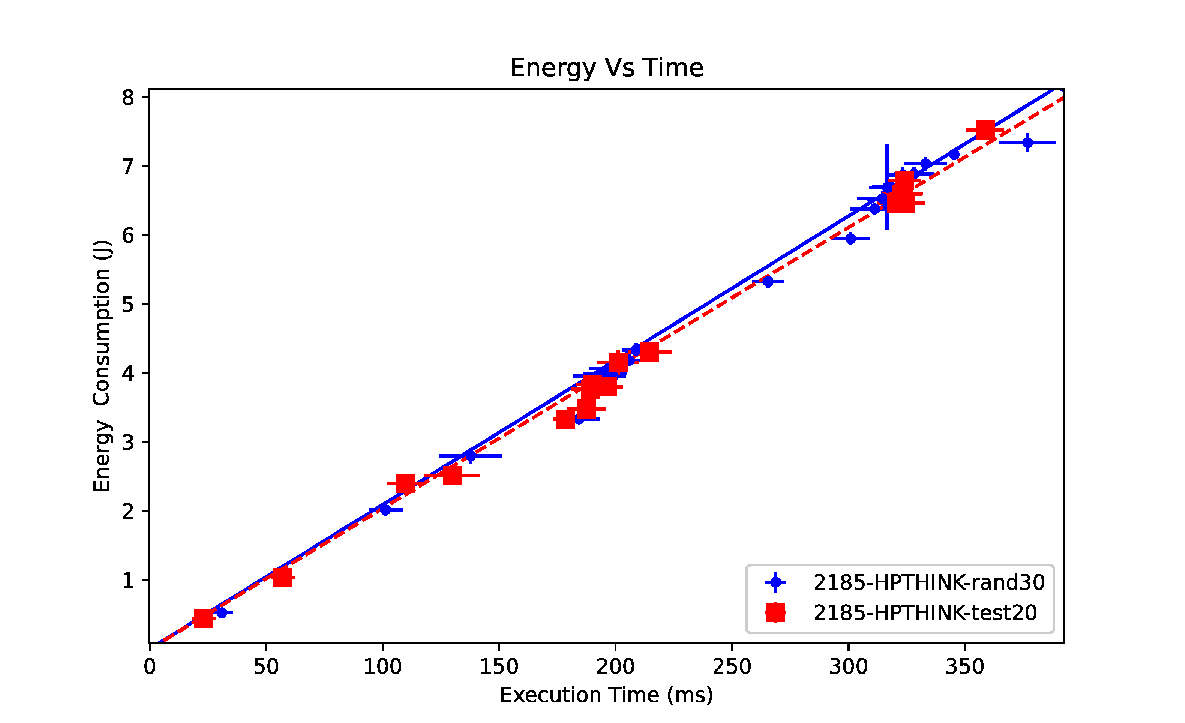}
    \\(b)
    \caption{Data and slope comparison between training set with 30 random solutions (Rand30 C++) and a) test set with 10 solutions and b) test set with 20 solutions for problem CSES 2185 in HPTHINK. Results improve with an increased test set.}
    \label{fig:2185}
  \end{center}
\end{figure}

Some of the bad results might be explained by a high number of outliers. Since each test contains only 10 solutions, the effect of some possible outliers is high. However, since the number of points is small, it is difficult even to detect those. We believe that if the training and test sets were bigger, the results would improve since this would reduce the influence of the outliers. For example, Figure \ref{fig:thinkrand30} b) shows that, for the HPTHINK machine, the worst problem was 2185 - Prime Multiples (column 8). The correct problem in the training set is only the $7^{th}$ nearest problem to the corresponding test, with a relative difference of 4.6\% between the test and training set. Figure \ref{fig:2185} a) shows the difference between the slope of the training set (blue) and the test set (red). It is possible to see several points, possible outliers, belonging to the test set placed below the red line.

 To see if we could improve the results of this specific problem (CSES 2185) by increasing the number of samples, we gathered 10 more solutions for the test set (leaving the training set the same). In this case, we improved the results such that the difference was reduced from 4.6\% to 2.9\% and the correct solution became the $5^{th}$ nearest problem. Figure \ref{fig:2185} b) shows the comparison of the slope of the increased training set with the test set. Now they are much closer, reducing outlier effects. Using the increased test set for the other machine, HPELITE, we could also reduce the error from 1.8\% to 0.5\%.  

Based on these results, we can give a partially positive answer to our third research question~\ref{rq:question3}:  \textit{Is it possible to identify what problem a set of solutions is aimed to solve based on their energy consumption profile?} We showed that there is some discrimination power in the energy consumption profile of a problem, making it possible to correctly classify them about 3.5 times better than by chance, or to reduce the search space by half. We also saw that the results might improve if we take larger sets for training and testing. This seems to indicate that this classification method deserves further investigation.

In the next subsection, we discuss some issues that may have affected our experiments, and thus our conclusions.

\subsection{Threats to Validity}
\label{subsec:threats}

There are some factors that might affect the quality or validity of our acquired data, analysis, and conclusions.

Although all the tests were performed in terminal mode, after restarting the system, we did not kill other background processes executed by the operating system, such as network, Xserver, cron, etc. These processes compete for CPU and could affect the measurements obtained. However, as the \texttt{time} command considers only the CPU time used by the C++ solution being measured, we believe that we have mitigated this issue.

\added{Overall, in our measurements with a single core enabled the discrepancy between
wall-clock and CPU time was below 10 ms. Although for some measurements we got a
larger difference, we consider that this is not a serious issue, since the value of
such difference is still small when compared to the total execution time.
}

RAPL reads Machine-Specific Registers (MSRs) once every 1 ms to update the energy measurements, which may \deleted{be RAPL may} lead to not-so-accurate results in case of short-running code
paths~\cite{hahnel2012measuring}. We tried to mitigate this effect by choosing programs that would run for at least 10 times this value. In our research, all the outliers are programs that ran at least for 1000 ms. However, the measurements of short-running programs may be affected by transient factors, such as scheduler decisions, Operating System services, and paging, among others. We tried to mitigate this issue by taking actions such as restarting the machines. Moreover, the energy consumption measurements were consistent, with low standard deviations.


In \cite{khan_rapl_2018} the authors state that there is a correlation between RAPL’s package power and CPU package temperature (at least for Haswell architecture). In our experiments, we did not cool down the CPU in between tests. We ran all the tests sequentially. So the rise of the temperature might have increased the measured energy consumption for programs that ran later in the tests when compared to the ones that ran at the beginning.

During the computations to detect outliers, we considered only the standard deviation of the mean energy consumed, ignoring the standard deviation of the mean execution time.  Considering this additional uncertainty might explain some of the outliers we detected.

\added{Our results were based only on the energy consumption reported by RAPL for the package domain, which corresponds to the energy consumption for the entire package of the CPU, including cores and uncore components, but does not include the energy consumption related to the RAM. We did not consider the RAM energy consumption because we could not measure it consistently in all machines using the same measurement approach. For example, by using the measurement framework \systime\, we were able to get the RAM energy consumption for machine HPTHINK, but not for machine HPELITE. Nevertheless, we could notice that the RAM energy consumption reported by RAPL was more relevant in case of the server machine XEON, where such value is usually around 10\% of the energy consumption.}

\section{Related Work}
\label{sec:related}

In the last decade, there has been an increased interest in the energy consumption of computer applications~\cite{pinto2017energyconcern,marantos2022bringing,koedijk2022energy,lima2019haskell}.

After its release, Intel’s Running Average Power Limit(RAPL)~\cite{weaver2012measuring}
has achieved much popularity, as it greatly facilitates the task of measuring
the energy efficiency of computer applications~\cite{khan_rapl_2018}. 
In more recent Intel processors, RAPL energy measurements are nearly
equal to plug power readings~\cite{khan_rapl_2018}.

Because of these reasons, RAPL has enabled the emergence of
several energy profiling tools on top of it ~\cite{kelley2016adaptive,beyer2020cpu,noureddine2022joular}, and it was used in many works focused on the energy efficiency of computational applications. Below we discuss some related work more closely related to ours.

Several works \cite{couto2017sblp,pereira2017energy,pereira2021ranking} tried to better understand the energy-impact of a programming language by using the solutions available in the Computer Language Benchmarks Game (CLBG).
These works measured the running time, energy consumption, and memory usage of solutions in different languages that aimed to solve the same task. Such works concluded that there is a very strong relationship between the running time of a program and its energy consumption and that total memory usage also influences the energy consumption of a program. 

Differently from these previous works, we focused on evaluating distinct solutions written in a single language. Another important difference is that, unlike CLBG, CSES solutions do not use either external C++ libraries or threads. Moreover, while the solutions of CSES, as well as the ones of CLBG, try to be as fast as possible, in the case of CSES we also explicitly mined C++ slower solutions, which perform dozens, or even hundreds, of times longer than the corresponding faster C++ solutions.

Despite these differences, our work came to similar conclusions. Specifically, when analyzing the outliers we could see that an outlier with an energy consumption higher than expected usually performs more allocations, while an outlier that consumes less energy than expected does the opposite. We think this is in consonance with~\cite{pereira2021ranking}, which suggests that may be more energy-efficient to allocate high amounts of memory at once.

When evaluating outliers, these previous works~\cite{couto2017sblp,pereira2017energy,pereira2021ranking} roughly compare the list of programs sorted by running time and by energy consumption and look for a mismatch. This approach may not be feasible when there are several programs with similar running time or energy consumption since a small deviation can produce a meaningful impact on the ordering. In our study, where we evaluated a single programming language, we think the usage of an approximation method, such as the least square method, can indicate more precisely the outliers. Moreover, it gives us a way to estimate the energy consumption of a program based on its execution time.

The work of Seng et al.~\cite{seng2014pentium4}, before the appearance of RAPL,
uses a power plug strategy to evaluate the effect of compiler optimizations
on the energy consumption of a Pentium 4. In their study, the authors of~\cite{seng2014pentium4}
used C++ solutions from the SPEC 2000 benchmark suite, and concluded
that when an optimization resulted in energy savings, this was
primarily due to a reduction in program execution time.
Our results, based on RAPL measurements, indicate that \deleted{this}
the conclusion remains valid in Pentium 5 machines.

The correct understanding of the meaning of a program is crucial for programming language processing~\cite{peng2021classification}, although it impossible to exactly determine
it~\cite{rice1953classes}. Recently, several approaches, based mainly on machine
learning techniques, tried to solve the algorithm classification problem~\cite{classification2023cgo}.
All these approaches take as input a representation derived from the source code of a
program (the LLVM intermediate representation of a program, for example) and try to
relate it to a given finite set of problems. Our classification approach has
a similar goal, but it does not depend on a program source code, it relies
only on the energy consumption profile produced by an executable program.
On the other hand, our approach classify problems for a particular machine configuration,
while the other approaches, which are based on some form of source code,
could provide a more machine-independent classification.

\section{Conclusions and Future Work}
\label{sec:conc}

The interest in the energy efficiency of software programs has been increasing. The running time of a program is the factor that greatly influences its energy consumption.

We evaluate the running time and the energy consumption of 900 C++ \added{and 450 Java} programs, aimed to solve 15 different tasks, mined from CSES, a popular programming contest site. We could confirm the general trend that faster programs consume less energy. Our test data presented a very high correlation between execution time and energy consumption. Moreover, we were able to use ordinary least squares to \replaced{fit a linear function}{predict}, with good precision, \replaced{that relates energy consumption of a program based on its execution time}{the energy consumption of a program based on its execution time}. A manual analysis of the programs with unexpected energy consumption revealed that \deleted{their memory usage is} often 
\added{they perform a different amount of allocation and deallocation operations}
\deleted{different} when compared to programs with similar execution times. 

\added{For our data set consisting of single-threaded programs, we showed that the availability of multiple cores may affect the energy consumption profile of Java solutions, while it has little impact in case of C++ solutions.}

The classification method based on the energy consumption profile of a problem showed good results, and we can conclude that it has some discriminatory power. We also provided evidence that the results might be improved with larger training and test sets.

All the data used in our analysis is publicly available in a git repository.

As future work, we intend to investigate further if it is possible to predict the energy consumption of C++ solutions, aimed at solving different problems, based on their running time. We would also want to perform a similar experiment to the one presented here for other programming languages used in programming competition sites, such as Python\deleted{ and Java},
and to check the relationship between energy consumption and the amount of CPU instructions executed. Moreover, we also want to do a more detailed investigation of the memory usage of programs and how it correlates with energy consumption.

Finally, we would like to further investigate the proposed classification method by gathering more data (increasing training and test sets) and using machine learning algorithms, such as k-nearest neighbors \cite{cover_nearest_1967}, to see if we can improve the results.  
We should also explore the usage of our approach to classify problems based on their energy consumption profile in different contexts, such as code obfuscation, and compare it with other classification approaches~\cite{classification2023cgo}.

\bibliographystyle{elsarticle-num} 
\bibliography{main}

\end{document}